\documentclass[11pt]{article}
\usepackage{amsmath,amssymb}
\usepackage{slashed}
\usepackage{xcolor} 
\usepackage{graphicx}
\usepackage{dcolumn} 
\usepackage{bm} 
\usepackage{epsfig}
\usepackage{epstopdf}
\usepackage{grffile}
\usepackage{color}
\usepackage{colordvi}
\usepackage{amsmath,amssymb}
\usepackage{rotating}
\usepackage{lscape}
\usepackage{cite}
\usepackage{float}
\usepackage{hyperref}
\usepackage{upgreek}
\usepackage{appendix}
\usepackage{multirow}

\newcommand{\keV}{{\rm\ keV}}
\newcommand{\GeV}{{\rm\ GeV}}

\def\comment#1{}

\newcommand{\beq}{\begin{eqnarray}}
	\newcommand{\eeq}{\end{eqnarray}}

\newlength{\myVSpace}
\def\beq{\begin{equation}}
	\def\eeq{\end{equation}}
\def\bea{\begin{eqnarray}}
	\def\eea{\end{eqnarray}}

\begin{document}
	\bibliographystyle{unsrt}
	\begin{titlepage}

		\begin{center}
			{\it }
			
		\end{center}
		\vskip 8pt
		
		\begin{center}
			
			\vspace*{15mm}
			\vspace{1cm}
			{\Large \bf Circular polarization of cosmic photons due to their interactions with  Sterile neutrino dark matter }
			
			\vspace{1cm}
			
			{\small \bf M. Haghighat \footnotemark[12], S. Mahmoudi\footnotemark[1], R.Mohammadi\footnotemark[34], S. Tizchang\footnotemark[5] and  S.-S. Xue\footnotemark[6] }
			
			\vspace*{0.5cm}

			$\footnotemark[1]$ {\it \small Physics Department, College of Sciences, Shiraz University 71454, Shiraz, Iran}\\
			$\footnotemark[2]$ {\it \small Islamic World Science Citation Center ISC, 71946-94173, Shiraz, Iran}\\
			$\footnotemark[3]$ {\it \small Iranian National Science and Technology Museum (INMOST), PO BOX: 11369-14611, Tehran, Iran }\\
			$\footnotemark[4]$ {\it \small School of Astronomy, Institute for Research in Fundamental Sciences (IPM), PO BOX 19395-5531, Tehran, Iran}\\
			$\footnotemark[5]$ {\it \small School of Particles and Accelerators, Institute for Research in Fundamental Sciences (IPM) PO BOX 19395-5531, Tehran, Iran}\\
			$\footnotemark[6]$ {\it \small ICRANet and Department of Physics, University of Rome``Sapienza''  P.le A. Moro 5, 00185 Rome, Italy}

			\vspace*{.2cm}
		\end{center}
		
		\vspace*{10mm}
		
		\begin{abstract}
			In this paper, we explore the possibility of the polarization conversion of a wide energy range of cosmic photons to the circular polarization through their interactions with  right handed  Sterile neutrinos as a candidate for dark matter. By considering the Sterile neutrino in the seesaw mechanism framework and right-handed current model, we examine the Faraday conversion $\Delta \phi_\text{\tiny{FC}}$ of gamma ray burst (GRB) photons at both the prompt and afterglow emission levels 
			as well as the radio photons emitted from our galaxy and extra-galactic sources  interacting with the Sterile neutrinos. Consequently, for the Sterile neutrino with mixing angle $\theta^2\lesssim 10^{-2}$ motivated by models with a hidden sector coupled to the sterile neutrino, the Faraday conversion can be estimated as $\Delta \phi_\text{\tiny{FC}}\lesssim 10^{-3}-10^{-18}$ rad for GRB,   $\Delta \phi_\text{\tiny{FC}}\lesssim 10^{-6}-10^{-11}$ rad for radio emission source from our galaxy and $\Delta \phi_\text{\tiny{FC}}\lesssim 10^{-6}-10^{-15}$ rad for extra-galactic sources. We also examine the V-mode power spectrum $C_{Vl}$ of the cosmic microwave background (CMB) at the last scattering surface. We show that the circular polarization power spectrum at the leading order is proportional to the linear polarization power spectrum $C_{pl}$ and the mixing angle where for $\theta^2\lesssim 10^{-2}$  leads to  $C_{Vl}\lesssim 0.01$  Nano-Kelvin squared.
			
		\end{abstract}
	\end{titlepage}

	\newpage
	\section{Introduction}
	Over the past century, the existence of dark matter (DM), the non-baryonic substance of the universe, which accounts for $26\%$ of the total energy density of the universe, has been discussed. The cosmological evidence  like curves in the galactic halos \cite{Ullio:2000bf} as well as astrophysical observations such as WMAP \cite{Page:2006hz} and Planck \cite{Ade:2015dga}, increase the DM existence probability. Besides cosmological and astrophysical evidence, it is crucial to attain information about interaction features of DM, if exist, with the standard model (SM) particles. Such information can be obtained from direct detection for instance in  XENON10 \cite{Angle:2011th} ,XENON100 \cite{Aprile:2011hi}, XMASS \cite{Abe:2008py}, CoGEANT \cite{Aalseth:2010vx}, DAMA \cite{{Bernabei:2008yi},{Belli:2011kw}}, PICASSO \cite{{Archambault:2009sm},{Archambault:2010jb}} and in indirect search using experiments such as production signatures at colliders \cite{Pani:2017qyd} or searching for annihilation and decay signals \cite{Bell:2017irk}. However, a different window into the nature of DM can be introduced in investigating of the circular polarization effects in scattering of the cosmic photons  from DM particles with various astrophysical sources.
	From the theoretical point of view, the circular polarization is generated from several mechanisms, mostly new physics interactions, which contribute to the Boltzmann equation. For example, forward scattering of the CMB photon from cosmic neutrino background (CNB) leads to the circular polarization of the CMB photon \cite{Mohammadi:2013dea}. CMB photons scattered from electrons can acquire circular polarization in the presence of  background fields such as Lorentz violation \cite{Bavarsad:2009hm},  magnetic field \cite{Cooray:2002nm, Bavarsad:2009hm}, non-commutative space-time  \cite{Batebi:2016ocb, Bavarsad:2009hm} and CP violation \cite{Bartolo:2019eac}. Furthermore, conversion of a linear to circular polarization for GRB photons in scattering from cosmic particles \cite{Batebi:2016efn} or production of circular polarization for the CMB from circularly polarized primordial gravitational waves \cite{Alexander:2018iwy} are also considered. \par
However, there are many sources for exploring the effects of the DM-photon scattering on the polarization production of  cosmic photons.  In addition to the CMB which provides a unique cosmological information at recombination epoch at the early universe, there are cosmic rays with a wide range of wavelengths
which can be used to study the properties of the DM particles.  For example, the GRBs as non-uniform pulses of gamma-ray radiation lasting commonly less than a minute, have detected at redshift less than ten \cite{Kumar:2014upa}. It is believed that they are produced at the end of massive star evolution and forming black holes \cite{Woosley:1993wj} or combining of compact objects \cite{Gehrels:2013xd}.  It can be seen at a random location on the sky and few times during a day. Generally, GRBs are followed-up by afterglow emissions including longer wavelength X-ray, optical, IR and radio frequencies \cite{Piran:2004ba}.  Meanwhile, the radio photons also can be considered through different sources such as  galactic supermassive black-hole inside the Milky Way, the distant radio galaxies  or from  the star formation in a way that by heating up the surrounding dust of a young star or exploding a massive young star as supernova after its born \cite{{Chambers},{Tabatabaei}}. 
	\par
	In theoretical term, among the SM particles only neutrinos can fulfill properties of a DM candidate. However, its small mass and large  coupling with the other SM particles keep neutrino relativistic at the epoch of freeze-out and it would only picture the hot DM \cite{Frenk:2012ph}. In the meantime, there are many models beyond the SM which provide one or more unknown particles with different masses, interaction, spin and strength to account for the DM (for instance see refs. \cite{{Lin:2019uvt},{Boyarsky:2018tvu}} and the references therein).  Nonetheless, in a large fraction of such models 
	 a weakly interacting spin 1/2 Majorana fermion is predicted which is singlet under the SM gauge group.  Also it can be found in the context of right-handed Sterile neutrino, for a review see for example \cite{Abazajian:2012ys}, and the right-handed current model see for example Refs.~\cite{{Xue:1997tz},{Xue:2016dpl}}. Furthermore, the sterile neutrino idea is powerful enough to explain the baryon asymmetry \cite{{Asaka:2005pn},{Shaposhnikov:2008pf}} and observed neutrino oscillations \cite{Boyarsky:2009ix} if it is considered as a triplet. With less  mass \cite{Asaka:2005an}, it can provide a viable DM through the seesaw mechanism \cite{Dodelson:1993je}. The seesaw mechanism is implemented in three tree level ideas so-called as type-I \cite{{Minkowski:1977sc},{Mohapatra:1979ia}}, type-II \cite{Cheng:1980qt} and type-III \cite{Foot:1988aq}.
	Nevertheless, there are some alternative extended models as well \cite{{Kusenko:2010ik},{Cox:2017rgn}}.
	\par
	Meanwhile, cosmological and astrophysical aspects of the massive Sterile neutrino  are studied extensively in literature \cite{{Canetti:2012kh}, {Bezrukov:2017ike}}.
	In this paper for the first time, we study the circular polarization production of the cosmic radiation  caused by the cosmic photons interacting with the Sterile neutrinos, as the Warm DM (WDM). This provides a new tool to explore the DM properties within  Type-I seesaw mechanism \cite{{Minkowski:1977sc},{Mohapatra:1979ia}} and the right-handed current model \cite{{Xue:1997tz},{Xue:2016dpl}}.\par
	This paper is organized as the following: we first present a brief review of the seesaw type I model and the right-handed current model in section \ref{sectionII}. In section \ref{sectionIII}, the time evolution of the Stokes parameters for photon-Sterile neutrino interaction is calculated by using the scalar mode perturbation of metric and the generation of circular polarization. The circular polarization arising from GRB-Sterile neutrino, radio foreground radiation-Sterile neutrino and CMB-Sterile neutrino forward scatterings are estimated in section \ref{sectionIV}. In section  \ref{sectionV} we give a summary and conclusion. Finally in Appendix \ref{appendix1} we give a  brief introduction on the polarized radiative transfer equation and its relation to the Faraday conversion  and Appendix \ref{appendix}, is devoted to the detail of calculation of the Boltzmann equation for the photon-Sterile neutrino interaction.
	\section{Right Handed Neutrinos}\label{sectionII}
	\subsection{Type-I seesaw}\label{subsectionII}
	Right-handed Sterile neutrinos are elegantly embedded in  the seesaw model. In type-I seesaw model the SM is extended by at least two heavy Sterile neutrino singlets $\nu_\text{\tiny{R}}^i$ ($i$ indicates the generation) with the following most general electroweak Lagrangian 
	\begin{equation}
		\mathcal{L}=\mathcal{L}_{\text{\tiny{SM}}}+y_{ij}^{\nu}\bar{\ell}_{\text{\tiny{L}}}^i\tilde{H} \nu_{\text{\tiny{R}}}^j +\frac{1}{2} M_\text{\tiny{R}}^i \bar{\nu_\text{\tiny{R}}^{i c}} \nu_\text{\tiny{R}}^{i}+h.c.,
		\label{seasawL}
	\end{equation}
	where $\mathcal{L}_{\text{\tiny{SM}}}$ denotes the electroweak Lagrangian of the SM  and $y_{ij}^{\nu}$ is a matrix
	of Yukawa interactions, $H$ is the Higgs
	doublet and $\tilde{H}=\epsilon\,H^{*}$, with $\epsilon$ is the anti-symmetric SU(2)-invariant tensor, $\ell_{\text{\tiny{L}}}=(\nu_{\text{\tiny{L}}}, e_{\text{\tiny{L}}})^{T}$ indicates the left handed lepton doublets and $\nu_\text{\tiny{R}}^{i c}=C\,\bar{\nu_\text{\tiny{R}}}^{i T}$ with $C=i\gamma_{2}\gamma_{0}$. Furthermore, $\nu^i_\text{\tiny{R}}$'s are SM gauge singlets, hence the Majorana mass term $M^i_\text{\tiny{R}}$ is allowed in addition to the Dirac mass $m_\text{\tiny{D}}$. Consequently, after the electroweak symmetry breaking one can obtain the Dirac mass as $m_\text{\tiny{D}}=y^\nu \left\langle H\right\rangle$ where by considering both Dirac and Majorana masses leads to the  neutrino mass matrix as follows
	\begin{eqnarray}\label{neutrinomassfull}
		\mathcal{M}_{\nu}=\left(
		\begin{tabular}{c c}
			$0$ & $m_\text{\tiny{D}}$\\
			$m_\text{\tiny{D}}^\text{\tiny{T}}$ & $M_\text{\tiny{R}}$
		\end{tabular}
		\right),
	\end{eqnarray}
	where $M_\text{\tiny{R}}$ and $m_\text{\tiny{D}}$ are $3\times3$ matrices. However,
	the eigenvalues of $M_\text{\tiny{R}}$ can be chosen to be at a scale much higher than the electroweak scale suppressing the Dirac mass term. Meanwhile, 
	to diagonalize the mass matrix, one needs a $6\times6$ mixing unitary matrix. In fact, the diagonalizing process occurs through two steps I) block diagonalizing and II) two unitary rotations. Therefore, there would be two sets of physical eigenstates related to the three light neutrinos of the SM particles. In the first set of eigenstates which are known as active neutrinos, the masses can be obtained as 
	\begin{equation}
		m_\nu=-m_\text{\tiny{D}}^\text{\tiny{T}}\,M_\text{\tiny{M}}^{\tiny{-1}}m_\text{\tiny{D}},
	\end{equation} 
	and the neutrinos belong to the $SU(2)$ doublets.  In the second set, one has a set of heavy right handed Majorana neutrinos which are gauge singlets with mass $M_\text{\tiny{M}}$ the eigenvalues of $M_\text{\tiny{R}}$.
	The scale of $M_\text{\tiny{M}}$ is not determined by experiment and different constraints are available from particle physics, astrophysics and cosmology with different consequences \cite{Drewes:2013gca}. 
	As a result, Sterile and the SM neutrinos mix with $\theta \equiv m_\text{\tiny{D}} M^{-1}_\text{\tiny{M}}$ mixing angle. Therefore,
	all of the Majorana  mass eigenstates can be represented by the flavor vector elements as:
	\begin{eqnarray}\label{masseig}
		N =V^\dagger_\text{\tiny{N}}\nu_\text{\tiny{R}}+\Theta^\text{\tiny{T}}\nu^c_\text{\tiny{L}}+h.c.\,, ~~~~\text{and}~~~~
		\upnu = V^\dagger_\nu \nu_\text{\tiny{L}}-U^\dagger_\nu \theta \nu_\text{\tiny{R}}^c+h.c.\,,
	\end{eqnarray}    
	where $V_\nu$ is the usual neutrino mixing matrix connecting the observed light mass eigenstates $\nu_i$ to the active flavor eigenstates:
	\begin{equation}
		V_\nu\equiv (1-\frac{1}{2}\theta\theta^\dagger)U_\nu,
	\end{equation}
	and $U_\nu$ is the unitary part of neutrino mixing matrix. Meanwhile, the corresponding parameters in the Sterile sector are $V_\text{\tiny{N}}$ and $U_\text{\tiny{N}}$ and the active-Sterile mixing  angle is 
	\begin{equation}
		\Theta\equiv \theta U_\text{\tiny{N}}^\star.
	\end{equation}
	Thus the Sterile neutrinos interacts with the SM particles  as follows 
	\begin{eqnarray}\label{SN}
		\mathcal{L}&\supset&\,\,\sum_{l}-\frac{g}{\sqrt{2}}\bar{N}\,\Theta^{\dagger}\,\gamma ^{\mu}\,l_{\text{\tiny{L}}}\,W_{\mu}^{+}\,-\sum_{l}\frac{g}{\sqrt{2}}\,\bar{l}_{\text{\tiny{L}}}\,\gamma^{\mu}\,\Theta\,N\,W_{\mu}^{-}-\frac{g}{2\cos \theta_{\text{\tiny{W}}}}\,\bar{N}\,\Theta^{\dagger}\,\gamma ^{\mu}\,\nu_{l_{\text{\tiny{L}}}}\,Z_{\mu}\,\nonumber\\
		&&-\frac{g}{2\cos \theta_{\text{\tiny{W}}}}\,\bar{\nu}_{l_{\text{\tiny{L}}}}\,\gamma^{\mu}\,\Theta\,N\,Z_{\mu}-\frac{g}{\sqrt{2}}\,\frac{M_{\text{\tiny{N}}}}{m_{\text{\tiny{W}}}}\,\Theta\,h\,\bar{\nu}_{l_{\text{\tiny{L}}}}\,N\,-\frac{g}{\sqrt{2}}\,\frac{M_{\text{\tiny{N}}}}{m_{\text{\tiny{W}}}}\,\Theta^{\dagger}\,h\,\bar{N}\,\nu_{l_{\text{\tiny{L}}}}.
	\end{eqnarray} 
	where $l=e, \mu, \tau$ and $\nu_l$ denotes the left handed SM neutrinos in the flavor eigenstates which can be expressed in terms of the mass eigenstates as $\nu_{l_{\text{\tiny{L}}}}=P_\text{\tiny{L}}(V_\nu \upnu + \Theta N)$. However, a neutral and massive Sterile neutrino depending on the galaxy phase space density, universal galaxy surface density and the DM density can be fit to a WDM scenario \cite{Dodelson:1993je}. Nevertheless, the Sterile neutrinos  can decay radiatively at loop level into the SM neutrinos as $N\rightarrow \nu_{l_{\text{\tiny{L}}}} + \gamma$. Furthermore, the dominant tree-level decay channel for the Sterile neutrino is $N\,\to\,\nu_{\alpha}\,\nu_{\beta}\,\bar{\nu}_{\beta}$ with the following total decay width \cite{{Pal:1981rm},{Barger:1995ty}}
	\begin{equation}\label{gamma}
		\Gamma\,=\,\frac{G_{\text{\tiny{F}}}^{2}\,M^{5}}{96\,\,\pi^{3}}\theta^2\,,
	\end{equation}
	where $\theta^2\equiv\sum_{i=e,\mu,\tau}|\theta_i|^{2}$ and $G_\text{\tiny{F}}$ is the weak Fermi constant. Therefore,
	by requiring the condition of Sterile neutrino  lifetime being longer than the age of the
	Universe $t_\text{\tiny{Universe}}=4.4\times10^{17}\,\text{sec}$ \cite{Ade:2015xua}, the mixing angle $\theta^{2}$  should be constrained as 
	\begin{equation}\label{theta}
		\theta^{2}\,<\,1\,\,\,(\frac{1\keV}{M})^5,
	\end{equation}
	where $M $ denotes the mass of the Sterile neutrino. Meanwhile, depending on models and considering the astrophysical constraints, one can find different bounds on the mixing angle From $\theta^2\ll 10^{-8}$ \cite{{Boyarsky:2018tvu},{Dolgov:2002wy}} to $\theta^2\leqslant 10^{-1}$ \cite{Bezrukov:2017ike}.
	However, there are also some  direct laboratory measurements with a weak bound on the mixing angle in the keV mass range as $\theta^2 \geqslant 10^{-4}$ \cite{{Holzschuh:1999vy},{Abdurashitov:2017kka}}.\\
	
	\subsection{Right handed effective coupling}
	
	Here, we would like to introduce the right-handed Dirac neutrinos as the DM candidates which can be coupled effectively to the SM particles through the right-handed current interactions with the SM intermediate gauge bosons \cite{{Xue:2016dpl},{Xue:2015wha},{Xue:1997tz},{Xue:1996fm}} such as 
	\begin{equation}\label{rhc}
		\mathcal{L}\supset g_\text{\tiny{R}} (g/\sqrt{2})~\bar {l}_\text{\tiny{R}}\gamma^\mu\nu_{l_{\text{\tiny{R}}}}W^-_\mu + h.c.\,,
	\end{equation}
	where $\bar{l}$ stands for a charged lepton.
	In fact, this model was motivated by the parity symmetry reconstruction at high energies without any extra gauge bosons. 
	The counterpart of (\ref{rhc}) in the quark sector has been also studied in Refs.\cite{{Xue:1996fm},{Rcurrent2017}}.\par
\begin{figure}
		\centering
		\includegraphics[scale=0.5]{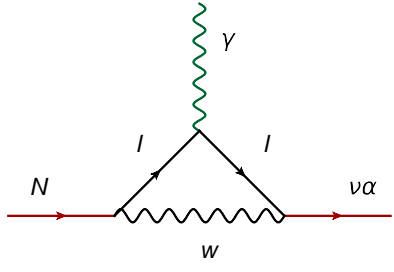}\\
		\caption{Radiative decay of right handed neutrino $ N\longrightarrow \nu_{a}+\gamma$ }\label{V}
	\end{figure}

 
	Besides the standard decay modes of $W$, according to the effective coupling of (\ref{rhc}), $W$ can also decay into the right handed fermions.  In the case of leptons, the partial width of the $W^{\pm}\rightarrow\,\bar{f_{i}}_{\text{\tiny{R}}}\,f_{j_{\text{\tiny{R}}}}$ decay mode  is determined by the following relation
	\begin{equation}
		\Gamma_{l_{ij}}\,=\,\frac{3\,g_{\text{\tiny{R}}}^{2}}{80\,\pi} \, m_{\text{\tiny{W}}}^{2}.
	\end{equation}
	This is while, the partial width of the $W^{\pm}\rightarrow\,\bar{f_{i}}_{\text{\tiny{R}}}\,f_{j_{\text{\tiny{R}}}}$ decay mode for quarks can be obtained as follows
	\begin{equation}
		\Gamma_{q_{ij}}\,=\,N_{c}\,|U_{ij}|^{2}\frac{3\,g_{\text{\tiny{R}}}^{2}}{80\,\pi} \, m_{\text{\tiny{W}}}^{2},
	\end{equation}
	where $N_{c}=3$ is the color factor and $U_{ij}$ is the element of CKM matrix.
	To get the total width of the $W$ gauge boson in this model, one must consider all leptons as well as quark decay modes. By considering all decay modes, the total width can be obtained as follows
	\begin{equation}
		\Gamma_{\text{\tiny{total}}}\,=\Gamma_{\text{\tiny{SM}}}+\delta\Gamma.
	\end{equation}
	where
	\begin{equation}
		\delta\Gamma\,\approx\,\frac{g_{\text{\tiny{R}}}^{2}}{4\,\pi} \, m_{\text{\tiny{W}}}^{2}.
	\end{equation}
	However, we should require that $\delta\Gamma$ does not exceed the experimental accuracy of $W$ decay width, $4.2\times 10^{-2} \text{GeV}$ \cite{Tanabashi:2018oca}. Therefore, the constraint on $g_{\text{\tiny{R}}}$ will be obtained as follows
	\begin{equation}\label{gr}
		g_{\text{\tiny{R}}}^{2}\leqslant 6\times10^{-3}.
	\end{equation}	
	Moreover,  life-time of the right handed neutrino as a DM candidate should be larger than the age of the universe. In fact, the right handed neutrinos can decay radiatively at loop level into the SM neutrinos as $ \nu_{\text{\tiny{R}}}\longrightarrow \nu_{a}+\gamma$, Fig (\ref{V}). Since in this model the radiative decay is a dominant  channel for decay of the right handed neutrinos, in the limit that $\nu_{a}$ is massless, the total decay width can be given as 
	\begin{equation}\label{decay width}
		\Gamma= \mathcal{G}^{2}\frac{M^{3}}{8\pi},
	\end{equation}
	where M is the mass of the right handed neutrino and the coupling constant $\mathcal{G}$ arises from the one-loop radiative corrections, Fig(\ref{V}), where after some calculation one has
	\begin{equation}\label{coupling}
		\mathcal{G} \approx \frac{\sqrt{2}}{2\pi^{2}}\,g_{\text{\tiny{R}}} G_{\text{\tiny{F}}}\,e\,m_{l}.
	\end{equation}
	Now by substituting (\ref{coupling}) into (\ref{decay width}), we arrive at the following relation
	\begin{equation}
		\Gamma(\nu_{\text{\tiny{R}}}\rightarrow \nu_{\text{\tiny{L}}} +\gamma)=\frac{\alpha}{4\pi^{4}}\,G_{\text{\tiny{F}}}^{2}\,\,m_{\ell}^{2}\,M^{3}\,g_\text{\tiny{R}}^{2}.
	\end{equation}
	Therefore, by requiring that  the right handed neutrino lifetime being longer than the age of the
	Universe $t_\text{\tiny{Universe}}=4.4\times10^{17}\,\text{sec}$ \cite{Ade:2015xua}, $g_{\text{\tiny{R}}}$  should be constrained as 
	\begin{equation}
		g_{\text{\tiny{R}}}^{2}\lesssim 10^{-2}\,(\frac{1.7\GeV}{m_{l}})^{2}\,(\frac{1\text{eV}}{\text{M}})^{3}.
	\end{equation}
	It should be noted that the interaction given in (\ref{rhc}) for the right handed neutrinos is very similar to  (\ref{SN}) for the Sterile neutrinos. Furthermore,  comparing (\ref{gr}) with the obtained  constraints on $\theta^2$ shows they are more or less in the same range.  Therefore, there is no difference between the scattering of cosmic photons from  the right handed neutrino or the Sterile one.  In fact, the obtained results in the next sections can be applied for both particles on the same footing.

	\section{Cosmic photons scattering from Sterile neutrino  }\label{sectionIII}
	The polarization of an ensemble of photons can be explained by the following density operator:
	\begin{equation}\label{stoke7}
		\hat{\rho}=\frac{1}{tr(\hat{\rho})}\int\frac{d^3p}{(2\pi)^3}\rho_{ij}({\bf p})\hat{D_{ij}}({\bf p}),
	\end{equation}
	where $\rho_{ij}$ shows the density matrix components in the phase space, ${\bf p}$ represents the momentum of cosmic photons and $\hat{D_{ij}}({\bf p})=a_i^\dagger({\bf p})a_j({\bf p})$ is the number operator of photons.
	This can also be decomposed into well-known Stokes parameters in the polarization space as follows
	\begin{equation}
		\hat{\rho}=\frac{1}{2}\left(
		\begin{matrix}
			I+Q &\,\, U-iV \\
			U+iV&\,\, I-Q \\
		\end{matrix}
		\right),\label{matrix}
	\end{equation}
	where $I$ is radiation intensity, $Q$ and $U$ represent linear polarization and circular polarization is given by the $V$ parameter. The $Q$ and $U$ quantities are influenced by orientation of coordinate system while $V$ and $I$ are coordinate independent. Therefore, a coordinate independent combination of $Q $ and $ U$ as $Q \pm i U$ is preferred.
	\par
	Stokes parameters for a propagating wave in the $\hat{z}$ direction are defined as
	\begin{eqnarray}
		I &\equiv& \langle E_x^2 \rangle + \langle E_y^2 \rangle \,\,\,\,\,\,\,\,\,\,\,\,\,\,\,\,\,\,\,\,\,\,\,\,\,\,\,\,\,\,\,\,\,\,\,\,\,\,\,\,\,\,\,\
		Q \equiv \langle E_x^2 \rangle - \langle E_y^2 \rangle \, , \nonumber \\
		U &\equiv& \langle 2E_xE_y \cos (\phi_x - \phi_y )\rangle \,\,\,\,\,\,\,\,\,\,\,\,\,\,\,\,\,\,\,\,\
		V\equiv \langle 2E_xE_y \sin (\phi_x - \phi_y )\rangle \, .
		\label{eqn:stokes}
	\end{eqnarray}
	The amplitudes and phases of waves in the $x$ and $y$ directions are defined with ($E_x$,$\phi_x$) and ($E_y$,$\phi_y$), respectively. The $\langle\cdot\cdot\cdot\rangle$ represents time averaging. In the standard model of cosmology, there is not  any physical mechanism to generate the $V$ parameter from the unpolarized cosmic photons. However, a linear polarization as is shown  in Appendix \ref{appendix1} can be converted to a circular one in a homogeneous medium through Faraday conversion (FC) defined as
	\begin{equation}
		\frac{\rm dV}{{\rm d} t}=  h\,U-g\,Q ,\label{faraday1}
		\end{equation}
		 where $g=\frac{d}{dt}\Delta \phi_\text{\tiny{FC}}|_{\tiny Q}$ and $h=\frac{d}{dt}\Delta \phi_\text{\tiny{FC}}|_{\tiny U}$ are the corresponding  Faraday conversion phase shifts caused by the conversion of linear polarization  $Q$ and $U$, respectively. It should be noted that in (\ref{faraday1}) based on the chosen reference frame, the Faraday conversion can be produced from a combination of $Q$ and $U$ parameters or either one of them. 
	 
	\par The time evolution of V-Stokes parameter or equivalently the component of density matrix can be obtained as \cite{Kosowsky:1994cy}
	\begin{equation}\label{stoke10}
		(2\pi)^3 \delta^3(0)(2p_0)\frac{d}{dt}\rho_{ij}({\bf p})=i\langle[H_I^0(t),D_{ij}^0({\bf p})]\rangle-\frac{1}{2}\int dt \langle[H_I^0(t),[H_I^0(0),D_{ij}^0({\bf p})]]\rangle,
	\end{equation}
	where the first order of the interaction Hamiltonian is given by $H_{I}^{0}$ and $p_0$ is the magnitude of photon momentum. The first term on the right-hand side is the forward scattering while the second one represents the higher-order collision terms.\par
		Therefore, by using the seesaw model and the right-handed coupling model, we can examine the effects of the photon-Sterile neutrino interaction on the polarization of the cosmic photons. To this end, we take Eqs. (\ref{SN}) and (\ref{stoke10}) into account to find the time evolution of the density matrix components  as follows ( see  the appendix \ref{appendix} for the detail of derivation):
	\begin{eqnarray}\label{rhodot}
		\frac{d}{dt}\rho_{ij}(p)\,&=&\,-\frac{\sqrt{2}}{12\,\pi\,p^{0}}\alpha\,\theta^{2}\,{G}_{\text{\tiny{F}}}\int d{\bf{q}}\,\, (\delta_{is}\rho_{s'j}(p)-\delta_{js'}\rho_{is}(p))\,f_\text{\tiny{DM}}({\bf{x}},{\bf{q}})\,
		\bar{u}_{r}(q)\,\,(1-\gamma^{5})\nonumber\\&&(q\cdot\epsilon_{s}\,\,\slashed{\epsilon}_{s^{'}}\,+\,q\cdot\epsilon_{s^{'}}\,\,\slashed{\epsilon}_{s})\,u_{r}(q)+\frac{\sqrt{2}}{24\,\pi\,p^{0}}\alpha\,\theta^{2}\,{G}_{\text{\tiny{F}}}\int d{\bf{q}}\,\, (\delta_{is}\rho_{s'j}(p)-\delta_{js'}\rho_{is}(p))\,\nonumber\\&&f_\text{\tiny{DM}}({\bf{x}},{\bf{q}})\,    \bar{u}_{r}(q)\,(1-\gamma^{5})\,\slashed{p}\,(\slashed{\epsilon}_{s^{'}}\,\slashed{\epsilon}_{s}\,-\,\slashed{\epsilon}_{s}\,\slashed{\epsilon_{s^{'}}})\,u_{r}(q).
	\end{eqnarray}
	
	Consequently, reconstruction of the Stokes parameters through the density matrix elements leads to the Boltzmann equations as follows
	\begin{equation}\label{DD25}
		\frac{d I}{dt}=C_{e\gamma}^I,
	\end{equation}
	\begin{equation}\label{B mode}
		\frac{d}{dt}\Delta_{P}^{\pm}=C_{e\gamma}^\pm \mp i\dot{\eta}^{\text{\tiny{P}}}_\text{\tiny{DM}}\Delta_{P}^{\pm}+\mathcal{O}(V),
	\end{equation}
	\begin{equation}\label{DDM27}
		\frac{d V}{dt}=C_{e\gamma}^V+\frac{1}{2}\,\big(\dot{\eta}_{\text{\tiny{DM}}}^{\text{\tiny{C}}-}\,\Delta_{P}^{+}\,+\,\dot{\eta}_{\text{\tiny{DM}}}^{\text{\tiny{C}}+}\,\Delta_{P}^{-}\big)\,,
	\end{equation}
	in which $\Delta_{P}^{\pm}\,=\,Q\pm\,i\,U$.  $C_{e\gamma}^I$, $C_{e\gamma}^V$ and $C_{e\gamma}^\pm$  demonstrate the contributions from the usual Compton scattering to the time evolution of $I$, $V$, and $\Delta_{P}^{\pm}$ parameters, respectively. Their explicit expressions are available in the literature for example see Refs \cite{{Zaldarriaga:1997ch},{Hu:1997hv}}.  Meanwhile, $\dot{\eta}^\text{\tiny{P}}_\text{\tiny{DM}}$ and $\dot{\eta}_\text{\tiny{DM}}^{\text{\tiny{C}}\pm}$ which are considered for the contribution of the photon-Sterile neutrino  scattering can be obtained as
	\begin{eqnarray}\label{Bmode1}
		\dot{\eta}^\text{\tiny{P}}_\text{\tiny{DM}}&=&\frac{\sqrt{2}}{3\pi p^{0}\,M}\,\,\alpha\,\,{G}_{\text{\tiny{F}}}\,\theta^{2}\,\int\,d{\bf{q}}\,f_\text{\tiny{DM}}({\bf x},{\bf q})\,\times (\varepsilon _{\mu\,\nu\,\rho\,\sigma}\epsilon_{2}^{\mu}\,\epsilon_{1}^{\nu}\,p^{\rho}\,q^{\sigma}),
	\end{eqnarray}
	and
	\begin{eqnarray}
		\dot{\eta}_\text{\tiny{DM}}^{\text{\tiny{C}}\pm}\,&=&\,\frac{\sqrt{2}}{3\pi p^{0}\,M}\,\,\alpha\,\,{G}_{\text{\tiny{F}}}\,\theta^{2}\,\int\,d{\bf{q}}\,f_\text{\tiny{DM}}({\bf x},{\bf q})\,\times \Big[(-\,q\cdot\epsilon_{1}\,q\cdot\epsilon_{2}-q\cdot\epsilon_{2}\,q\cdot\epsilon_{1})\nonumber\\
		&&\pm i (q\cdot \epsilon_{1}\,q\cdot\epsilon_{1}\,-q\cdot\epsilon_{2}\,q\cdot\epsilon_{2})\Big],
	\end{eqnarray}
	where the incoming photons can be chosen from a wide range of low to high energy cosmic photons.
	\section{V-mode and circular polarization of cosmic photons }\label{sectionIV}
	Astrophysical searches of WDM candidate in new physics are essential part of the experimental efforts to explore the nature of WDM. In this section, we propose an indirect method to search for WDM via studying WDM effects on cosmic photons polarization through cosmic photon-sterile neutrino forward scattering. The strategy is to search for WDM signals in the regions of sky with the highest expectation for WDM  aggregation. Of these regions, the center of Galactic is one of the most promising locations for WDM searches and polarimetry of cosmic rays which come from these regions. Therefore, we consider GRB photons, galactic and extra galactic radio photons and the CMB photons as sources of cosmic rays to calculate the amount of possible circular polarization for each case via their scattering from the Sterile neutrino. This study might open a new observational window to explore the nature of DM. 
	\subsection{  Circular polarization of the GRB photons }
  Several reports for the polarization measurement for the GRBs have been prepared during recent years. For instance, the linear polarization of the prompt emission from GRB 021206 has been reported at the level of $80\%$\cite{Coburn:2003pi}. 
		The GRB polarimeter (GAP) has observed  $70\%$ and $84\%$ degrees of linear polarization of  GRB 110301A and GRB 110721A, respectively \cite{Yonetoku:2012wz}. Furthermore, the linear and circular polarization at afterglow radiation in GRB 121024A have been reported of the order of $28\%$ and $0.6\%$, respectively \cite{Wiersema:2014bha}. However, the linear polarization of GRB is mostly originated from synchrotron emission \cite{Troja}. Meanwhile, the GRB circular polarization  can be generated either due to a large-scale ordered magnetic field or Faraday conversion at the late time of the GRB radiation \cite{Bavarsad:2009hm, Cooray:2002nm, Batebi:2016ocb, Batebi:2016efn}.  \par
	In this subsection, we will consider the GRB photon-sterile neutrino forward scattering and 
	estimate the Faraday conversion phase shift in two cases: (i) GRB photons at the  prompt emission interacting with Sterile neutrinos passing through internal and external shocks, (ii)  GRB photons at the afterglow intermediate emission interacting with Sterile neutrinos on the way of their propagation.
	\par
	Based on the time evolution of the V-Stokes parameter in (\ref{DDM27})
	and considering (\ref{faraday}), the Faraday conversion  of the scattered photons from the  Sterile neutrinos  evolve as follows
	\begin{equation}\label{deltaphi}
		\Delta \phi_\text{\tiny{FC}}|_\text{GRB}\,=\,\frac{\sqrt{2}}{6\pi}\,\alpha\,G_\text{\tiny{F}}\,\theta^{2}\,\int \frac{dt}{p_0}\,f_\text{\tiny{DM}}({\bf{x}})\,v^{2}_\text{\tiny{DM}}\,(\hat{v}_{\alpha}\,\hat{q}_{\beta}\,\epsilon_{1}^{\alpha}\,\epsilon_{1}^{\beta}-\hat{v}_{\alpha}\,\hat{q}_{\beta}\,\epsilon_{2}^{\alpha}\,\epsilon_{2}^{\beta}),
	\end{equation}
	where $f_\text{\tiny{DM}}({\bf{x}})$ is proportional to the mass density of WDM as  $\rho_\text{\tiny{DM}}({\bf{x}})=M f_\text{\tiny{DM}}({\bf{x}})$ and $p_0=|{\bf p}|$.  \par
	There are two possibilities for the forward scattering of GRB photons from Sterile neutrino: i) At the prompt emission level, ii) At the afterglow intermediate interaction level. The former is strongly model dependent and the polarization of high energy GRBs can be produced by scattering from WDM located close to the GRB prompt emission location.
	Meanwhile, the latter is model independent as the polarization of afterglow GRB photons are affected by scattering from Sterile neutrinos  on their way to reach the detector.  \par
	We first estimate the Faraday conversion caused by the prompt emission-Sterile neutrino scattering. We consider a simple model in which the Sterile neutrino WDM can abundantly be produced in  supernovae cores.  Therefore, by scattering off the local Sterile neutrino, 
	the linear polarization of the GRB photons at the prompt emission level
	can be converted to the circular polarization as 
	\begin{equation}\label{deltaphi1}
		\Delta \phi_\text{\tiny{FC}}|_\text{GRB}\,=\,\,10^{-32}\theta^{2}\,(1+z)^{2}\,\,\,\,\,(\frac{\GeV}{p_{0}})(\frac{\rho_\text{\tiny{DM}}}{10^{-41}\GeV^{4}})\,(\frac{v_\text{\tiny{DM}}}{10^{-3}})^2\int \frac{dl}{10^{10}\text{cm}}\,(\hat{v}_{\alpha}\,\hat{v}_{\beta}\,\epsilon_{1}^{\alpha}\,\epsilon_{1}^{\beta}-\hat{v}_{\alpha}\,\hat{v}_{\beta}\,\epsilon_{2}^{\alpha}\,\epsilon_{2}^{\beta}),
	\end{equation}
	where the mass density of the Sterile neutrinos is assumed to be $\rho_\text{\tiny{DM}}\sim 10^{-41}\GeV^{4} $ \cite{Guver:2012ba}  
	and we have supposed that the Prompt $\gamma$-ray emission occurs at distance $\sim 10^{10}$ cm from the center.\par
	However, in the second case, the  afterglow radiation caused by the GRB photons can interact with the Sterile neutrinos in its way to the earth. Therefore, the GRB linear polarization is expected to be suppressed by the Faraday conversion phase shift.
	The integration over time in (\ref{deltaphi}) can alternatively convert to integration over the redshift as $\int_{t'}^0 dt=\int_0^{z'} dz/[(1+z)H(z)]$ with $H(z)=H_0[(\Omega_r(1+z)^4+\Omega_{M}(1+z)^{3}+\Omega_{\Lambda}]$ where  $\Omega_r\simeq 10^{-4}$, $\Omega_M\simeq 0.3$ and $\Omega_\Lambda\simeq 0.7$ are the present densities of radiation, matter plus DM and dark energy, respectively, and  $H_0=67.4\text{km/s/Mpc}$ is the value of  Hobble constant at the present time \cite{Aghanim:2018eyx}. 
	Meanwhile, the bulk velocity of WDM, number density and energy of contributing particles  depend on the redshift \cite{{Hu:1999vq},{Vahedi:2018abn}} as
	\begin{equation}
		v_\text{\tiny{DM}}\,=\,v_\text{\tiny{0DM}}\,(1+z)^{-1/2},\,\,\,\,\,\,\,\,\, p_{0}=p_{0,0} (1+z),\,\,\,\,\,\,\,\,\,\ \rho_\text{\tiny{DM}}=\rho_\text{\tiny{DM0}}(1+z)^3,
	\end{equation}
	where $v_\text{\tiny{0DM}}$, $p_{0,0}$ and $n_\text{\tiny{0DM}}$ are obtained at the present time. \par 
	By taking (\ref{deltaphi}) into account and integrating over the red shift $z$, the Faraday conversion can be estimated as follows
	\begin{equation}\label{delta}
		\Delta \phi_\text{\tiny{FC}}|_\text{GRB}\,=\,10^{-14}\theta^{2}\,\,(\frac{\rm\ keV}{p_0})\,\,\,(\frac{\rho_\text{\tiny{DM}}}{10^{-47}\GeV^{4}})\,(\frac{v_\text{\tiny{DM}}}{10^{-3}})^2,
	\end{equation}
	where we have supposed that the Sterile neutrino  has the same global mass density of DM today $\rho_\text{\tiny{DM}}\simeq 10^{-47}\GeV^{4}$\cite{Aghanim:2018eyx}.
	Conventionally, one can rewrite the above equation as a function of the decay rate  and the mass of Sterile neutrino (see (\ref{gamma})). Therefore, the Faraday conversion due to the GRB photon-Sterile neutrino scattering depends on the mass of the Sterile neutrino as follows 
	\begin{equation}
		\Delta \phi_\text{\tiny{FC}}|_\text{GRB}= 10^{-14}(\frac{\rm\ keV}{M})^5(\frac{\rm\ keV}{p_0}) \,(\frac{\rho_\text{\tiny{DM}}}{10^{-47}\GeV^{4}})\,(\frac{v_\text{\tiny{DM}}}{10^{-3}})^2,
	\end{equation}
	where the life time of the Sterile neutrino  $\tau_\text{\tiny{N}}= \tau_\text{\tiny{DM}}\sim 10 \times$ universe life time.
	\begin{table}
		\centering    \caption{GRB Faraday conversion phase shift due to photon-Sterile neutrino DM interaction for the electromagnetic spectrum, regarding $z=1$ and $\rho_\text{\tiny{DM}}=10^{-47}\GeV^{4}$\cite{Aghanim:2018eyx}}
		\label{tabel1}
		\scalebox{1}{
			\begin{tabular}{ l|l l l l }
				\hline
				& GRB types & $\lambda$ (cm)$\approx$ & $\,\,\,\,\,\,\Delta\,\phi_\text{\tiny{FC}}|_{\theta^{2}\approx10^{-2}-10^{-6}}\simeq$\\ \hline\hline
				\multirow{1}{*}{Prompt emission} & Prompt & $ 10^{-13}$& \,\,\,\,\,\,$\,10^{-21}-10^{-25}$ \\
				
				\hline
				\multirow{7}{*}{Afterglow emission} &$\gamma$ ray & $ 10^{-10}$& \,\,\,\,\,\,$\,10^{-18}-10^{-22}$ \\
				& X ray &$ 10^{-8}$& \,\,\,\,\,\,$\,10^{-16}-10^{-20}$\\
				&UV &$ 10^{-6}$ & \,\,\,\,\,\,$\,10^{-14}-10^{-18}$\\
				&Visible &$ 10^{-4}$ &\,\,\,\,\,\,$\,10^{-12}-10^{-16}$  \\ 
				& Infrared &$ 10^{-3}$&\,\,\,\,\,\, $\,10^{-11}-10^{-15}$\\
				&Microwave& $ 1$ &\,\,\,\,\,\,$\,10^{-8}-10^{-12}$ \\
				&Radio& $10^{5}$ &\,\,\,\,\,\,$\,10^{-3}-10^{-7}$ \\\hline
			\end{tabular}
		}
	\end{table}
	However, the Faraday conversion for the GRB afterglow spectrum scattering from the Sterile neutrinos can be estimated from (\ref{delta}) for different mixing angles in the range $\theta^{2}\approx 10^{-2}-10^{-6}$ as are shown in Tab.\ref{tabel1}.
		\subsection{Circular polarization of Galactic and Extragalactic radio sources}
		The radio synchrotron radiation is emitted from astronomical radio sources.  The astronomical radio sources are objects in our galaxy (the Milky Way) and extragalactic. Some of the famous candidates for the radio sources at the center of Milky Way, are galactic supermassive black hole Sagittarius A* \cite{Munoz:2011ie} and Supernova remnants such as Cassiopeia A \cite{Fesen:2006zma} and Crab Nebula   
		\cite{Ritacco:2018dsu}. The Centaurus A \cite{Zemcov:2009zu}, Blazar S5 0716+71\cite{Cenacchi:2009tg}, Messier 87 (M87) \cite{Kuo:2014pqa} and Messier 81* (M81*)\cite{Brunthaler:2006ch} radio galaxies are some of the notable sources of extra-galactic radio sources. \\
		Due to the magnetic field properties of radio galaxies, radio emission should highly be polarized. The most likely process which gives rise to linear and circular polarization of some radio sources in galactic and extragalactic nuclei is Faraday rotation and conversion \cite{{Tucci:2012qt},{Gruzinov:2019gda}}.
		In general, observation suggest the linearly polarization  at the degree of $\sim 10\%$ mainly due to the synchrotron radiation from relativistic electrons. However, the circular polarization is less than $0.5 \%$ \cite{Beckert:2001az, ERS, Brunthaler:2006ch, Brunthaler:2001dq, Mesa:2002km, Bower:1999vm, Bower:2002xq, crab, crab2, crab3}. Also the circular polarization in the absence of a linear polarization  has been detected as well \cite{Brunthaler:2006ch, Brunthaler:2001dq, Mesa:2002km}. \\ 
		Similar to the GRB photons, we  estimate the Faraday conversion due to the radio radiation-Sterile neutrino interaction for the intergalactic sources as follows
		\begin{eqnarray}\label{Sgr}
			\Delta \phi_\text{\tiny{FC}}|_{\text{\tiny{Sgr A*}}}\,&\approx&\,\,10^{-4}\theta^{2}\,(1+z)^{2}\,\,\,\,\,(\frac{6.5 \times 10^{-7}\text{eV}}{p_{0}})(\frac{\rho_\text{\tiny{DM}}}{10^{-41}\GeV^{4}})\,(\frac{v_\text{\tiny{DM}}}{10^{-3}})^2\nonumber\\
			&&\int \frac{dl}{2.4\times10^{22}\text{cm}}\,(\hat{v}_{\alpha}\,\hat{v}_{\beta}\,\epsilon_{1}^{\alpha}\,\epsilon_{1}^{\beta}-\hat{v}_{\alpha}\,\hat{v}_{\beta}\,\epsilon_{2}^{\alpha}\,\epsilon_{2}^{\beta}),
		\end{eqnarray}
		where $l$ demonstrates the distance of  intergalactic radio sources from the Earth.
		Using  (\ref{Sgr}) one can find  the values of the Faraday conversion for some intergalactic radio sources with $z\ll 1 $ as are listed in the first three rows of Tab. \ref{tabel2}. Meanwhile, the Faraday conversion due to the forward scattering of radio radiation-Sterile neutrino for the  radio galaxy M81*  source can be estimated as
		\begin{eqnarray}\label{M81}
			&&\Delta \phi_\text{\tiny{FC}}|_{\text{\tiny{M81*}}}\,\approx \Big(10^{-4}\theta^{2}\,(1+z)^{2}\,\,\,(\frac{\rho_\text{\tiny{DM}}}{10^{-41}\GeV^{4}})\int \frac{dl}{4\times10^{22}\text{cm}}\,(\hat{v}_{\alpha}\,\hat{v}_{\beta}\,\epsilon_{1}^{\alpha}\,\epsilon_{1}^{\beta}-\hat{v}_{\alpha}\,\hat{v}_{\beta}\,\epsilon_{2}^{\alpha}\,\epsilon_{2}^{\beta})\nonumber\\
			&& +5.35\times 10^{-8}\theta^{2}\,(1+z)^{2}\,\,\,(\frac{\rho_\text{\tiny{DM}}}{10^{-47}\GeV^{4}})\,\int \frac{dl}{1.096\times10^{25}\text{cm}}\,(\hat{v}_{\alpha}\,\hat{v}_{\beta}\,\epsilon_{1}^{\alpha}\,\epsilon_{1}^{\beta}-\hat{v}_{\alpha}\,\hat{v}_{\beta}\,\epsilon_{2}^{\alpha}\,\epsilon_{2}^{\beta}) \Big)\nonumber\\
			&&\,\,\,\,\,\,\,\,\,\,\,\,\,\,\,\,\,\,\,\,\,\,\,\,\,\,\,\,\,\,\,\,\,\,\,\,\,\,\,\,\,\,(\frac{6.5 \times 10^{-7}\text{eV}}{p_{0}})\,\,(\frac{v_\text{\tiny{DM}}}{10^{-3}})^2,\nonumber\\
		\end{eqnarray}
		where the first term corresponds to the  Faraday conversion effect inside the radio galaxy and the second one represents the Faraday conversion arising from photon traveling out of the radio galaxy toward  the Earth. For simplicity, the DM mass density in other galaxies is considered at the same order of magnitude of the DM mass density in our galaxies. Nevertheless, the DM mass density is considered as the local DM mass density $\rho_\text{\tiny{DM}}= 10^{-41}\GeV^4$ inside the galaxies and for the outside of galaxies it has been taken to be equal to the global mass density  $10^{-47}\GeV^4$. As the second term in (\ref{M81}) in comparison with the first one  shows, due to the value of the DM mass density, the effect of the Faraday conversion inside the radio galaxies is the dominant one.  However, the Faraday conversion for different radio galaxies are given the Tab.\ref{tabel2}.. It should be noted that for the extra-galactic sources, DM mass density can be considered as a free parameter as well as the Sterile neutrino-neutrino mixing angle. Therefore, measurement on the Faraday conversion can open a new window to probe the DM mass density in radio galaxies.
		\begin{table}
			\centering    \caption{Faraday conversion phase shift due to the interaction of Sterile neutrinos with photons originated from Galactic and Extra galactic radio sources.}
			\label{tabel2}
			\scalebox{1}{
				\begin{tabular}{ l|l l l l }
					\hline
					& Radio source & $\lambda$ (cm)$\approx$ & $\,\,\,\,\,\,\,\,l$ (ly)$\approx$& $\,\,\,\,\,\,\Delta\,\phi_\text{\tiny{FC}}|_{\theta^{2}\approx10^{-2}-10^{-6}}\simeq$\\ \hline\hline
					\multirow{3}{*}{Inter-galactic} & Sagittarius A*\cite{Bower:2002xq,Jimenez-Rosales:2018mpc} & \,\,\,\,\,$ 200$&\,\,\,\,\,\,\,\,\,$26000$ &\,\,\,\,\,\,\,\,\,\,\,\,\,\,\,\,\,\,$10^{-6}-10^{-10}$ \\
					& Cassiopeia A \cite{Fesen:2006zma, cassiopia}&$\,\,\,\,\, 600$&\,\,\,\,\,\,\,\, $11000 $&\,\,\,\,\,\,\,\,\,\,\,\,\,\,\,\,\, $10^{-6}-10^{-10}$\\
					& Crab Nebula \cite{crab,Kaplan:2008qm}& $\,\,\,\,\, 400$ &\,\,\,\,\,\,\,\,\,\,$6500$&\,\,\,\,\,\,\,\,\,\,\,\,\,\,\,\,\,\,$\,10^{-7}-10^{-11}$ \\
					\hline
					\multirow{4}{*}{Extra-galactic} & M81* \cite{Brunthaler:2001dq,Ros:2011nr}&\,\,\,\, $ 200$&\,\,\,\,\,\,\,\,\,$11\times10^{6}$ &\,\,\,\,\,\,\,\,\,\,\,\,\,\,\,\,\,\,$\,10^{-6}-10^{-10}$ \\
					& M87 \cite{Kuo:2014pqa,Bird:2010rd}& \,\,\,\,\,\,$ 0.1$&\,\,\,\,\,\,\,\, $53\times 10^{6}$&\,\,\,\,\,\,\,\,\,\,\,\,\,\,\,\,\,\,$\,10^{-11}-10^{-15}$\\
					&Centaurus A \cite{Zemcov:2009zu,Majaess:2010gr}&\,\,\,\,\, $ 1$ &\,\,\,\,\,\,\,\,\,$13\times 10^{6}$  &\,\,\,\,\,\,\,\,\,\,\,\,\,\,\,\,\,\,$\,10^{-8}-10^{-12}$\\
					&Blazar S5 0716+71 \cite{Cenacchi:2009tg, Larionov:2013mna}&\,\,\,\,\,$ 200$ &\,\,\,\,\,\,\,\, $3.6\times10^{9}$&\,\,\,\,\,\,\,\,\,\,\,\,\,\,\,\,\,\,$\,10^{-7}-10^{-11}$  \\ \hline
				\end{tabular}
			}
		\end{table}
	
	\subsection{Circular polarization of the CMB photons}
	In this section, we discuss about the circular polarization of the CMB photons in the conformal time $\eta$ due to photon-Sterile neutrino  scattering. 
	To this end, we focus on the left-hand side of the Boltzmann equation (\ref{DDM27}) including the information of photon propagation in the flat Friedman-Robertson-Walker(FRW) background space-time. As the circular polarization in presence of the scalar perturbation is dominant comparing to the vector and tensor perturbation, only the scalar perturbation is added to the metric and we neglect the vector and tensor perturbations.
	However, the DM distribution function is indicated as \cite{ {MC:2003},{Ma:1995ey},{Lesgourgues:2006nd}}:
	\begin{equation}\label{Distribution}
		f_\text{\tiny{DM}}(\vec{\mathbf{x}},\vec{\mathbf{q}},\eta)=f_\text{\tiny{DM0}}[1+\Psi(\vec{\mathbf{x}},\vec{\mathbf{q}},\eta)],
	\end{equation} 
	where  $f_\text{\tiny{DM0}}(\vec{\mathbf{x}},\vec{\mathbf{q}},\eta)$ shows the
	zeroth-order distribution, $\Psi(\vec{\mathbf{x}},\vec{\mathbf{q}},\eta)$ is the  perturbed part and $\vec{\mathbf{q}}=q \hat{\mathbf{n'}}$ where $\hat{\mathbf{n'}}$ indicates direction of the DM velocity.
	Neglecting the collision term on the right hand side of the Boltzmann equation, the phase space distribution of the Sterile neutrino  can be obtain as follows
	\begin{equation}\label{bonui}
		\frac{\partial f_\text{\tiny{DM}}}{\partial \eta}+i \frac{q}{ \varepsilon_\text{\tiny{DM}}}(\vec{\mathbf{K}}\cdot\hat{\mathbf{n'}})\Psi+
		\frac{d\ln f_\text{\tiny{DM0}}}{d\ln q}[\dot\varphi-i \frac{ \varepsilon_\text{\tiny{DM}}}{q}(\vec{\mathbf{K}}\cdot\hat{\mathbf{n'}})\psi]=0,
	\end{equation} 
	where $\varphi$ and $\psi$ indicate the scalar metric perturbation in the Newtonian gauge \cite{Mukhanov:1992}, $\vec{\mathbf{K}}$ is wave number of the Fourier modes of the scalar perturbations and $\varepsilon_\text{\tiny{DM}}=(q^2+a(\eta)^2\,M^2)^{1/2}$ with the scale factor $a(\eta)$. Meanwhile, the angular
	dependence of the perturbation can be expanded in a series of Legendre polynomials $P_l(\mu')$ as follows
	\begin{equation}\label{App1}
		\Psi(\vec{\mathbf{K}},q,\mu',\eta)=\sum_{l=0} (-i)^l(2l+1)\Psi_{ l}(\vec{\mathbf{K}},\eta)P_l(\mu'),
	\end{equation}
	with $\mu'= \hat{\mathbf{K}}.\hat{\bf n'}$.
	Now, we expand (\ref{DDM27}) in terms of $\Psi_{ l}$ and $\mu'$ as follows
	\begin{eqnarray} \label{taupm}
		\dot{V}&\simeq&\frac{\sqrt{2}}{3\pi\,p_{0}}\,\alpha\,\theta^{2}\,G_\text{\tiny{F}}\,\left[(\eta_\text{\tiny{B}}-i \,\eta_\text{\tiny{A}})\,\Delta_{P}^{+\,(S)}\,+\,(\eta_\text{\tiny{B}}+i \,\eta_\text{\tiny{A}})\,\Delta_{P}^{-\,(S)}\right]\\ \nonumber 
		&\times& \frac{4\pi}{3}(\frac{1}{(2\pi)^{3}})\int q^2\, dq\,\frac{q^2}{\varepsilon_\text{\tiny{DM}}}\, f_\text{\tiny{DM0}}[\Psi_0-2\Psi_2]\ \nonumber\\
		&\simeq&\dot{\eta}_\text{\tiny{DM}}^\text{\tiny{C-}}\,\Delta_{P}^{+}\,+\,\dot{\eta}_\text{\tiny{DM}}^\text{\tiny{C+}}\,\Delta_{P}^{-},
	\end{eqnarray}
	where
	\begin{eqnarray}
		\dot{\eta}_\text{\tiny{DM}}^{\text{\tiny{C}}\pm}\,&=&\,\dot{\eta}_\text{\tiny{DM}}^{\text{\tiny{C}}}\,(\eta_\text{\tiny{B}}\,\pm\,i\,\eta_\text{\tiny{A}}),    \,\,\,\,\,\,\,
	\end{eqnarray}
	with
	\begin{eqnarray}
		\eta_\text{\tiny{A}}\,&=&\,\left(\hat{\mathbf{K}}\cdot\mathbf{\epsilon}_{1}\right)^{2}-\left(\hat{\mathbf{K}}\cdot\mathbf{\epsilon}_{2}\right)^2\,,\,\,\,\,\,\,\,\,\,\,
		\eta_\text{\tiny{B}}\,=-2\hat{\mathbf{K}}\cdot\mathbf{\epsilon}_{1}\,\,\hat{\mathbf{K}}\cdot\mathbf{\epsilon}_2,
	\end{eqnarray}
	and
	\begin{equation} \dot{\eta}_\text{\tiny{DM}}^{\text{\tiny{C}}}\,=\,\frac{\sqrt{2}}{3\pi\,p_{0}}\,\alpha\,\theta^{2}\,G_{\text{\tiny{F}}}\,\frac{1}{(2\pi)^{3}}\left[\delta p_\text{\tiny{DM}}-(\bar\rho_\text{\tiny{DM}} +\bar{P}_\text{\tiny{DM}})\sigma_\text{\tiny{DM}}\right],
	\end{equation}
	in which $\sigma_\text{\tiny{DM}}$ is shear stress  and $\bar{\rho}_\text{\tiny{DM}}$ and $\bar{P}_\text{\tiny{DM}}$ are the unperturbed energy densities and pressure defined as \cite{Ma:1995ey}
	\begin{equation}\label{rohbar}
		\bar\rho_\text{\tiny{DM}} =\, a^{-4}\int\, q^2\, dq\, d\Omega\,\varepsilon_\text{\tiny{DM}}\,f_\text{\tiny{DM0}},\,\,\,\,\,\,\,\,\,\bar{P}_\text{\tiny{DM}} =\frac{1}{3}\, a^{-4}\int\, q^2\, dq\, d\Omega\,\frac{q^2}{\varepsilon_\text{\tiny{DM}}}\,f_\text{\tiny{DM0}},
	\end{equation}
	\begin{equation}
		(\bar\rho_\text{\tiny{DM0}}+\bar P_\text{\tiny{DM0}})\sigma_\text{\tiny{DM}} =\frac{8\pi}{3} \,a^{-4}\int\, q^2\, dq\, d\Omega\,\frac{q^2}{\varepsilon_\text{\tiny{DM}}}\,f_\text{\tiny{DM0}}\Psi_2,
	\end{equation}
	and $\delta P_\text{\tiny{DM}}$ as the perturbation of pressure is  \cite{Ma:1995ey}
	\begin{eqnarray}
		\delta P_\text{\tiny{DM}} =\frac{4\pi}{3} \,a^{-4}\int\, q^2\, dq\,\frac{q^2}{\varepsilon_\text{\tiny{DM}}}f_\text{\tiny{DM0}}\Psi_0,
	\end{eqnarray}
	with
	\begin{eqnarray}
		\dot{\Psi}_0 &=& -\frac{q{\bf K}}{\varepsilon}\Psi_1-\dot{\phi}\frac{d\ln f_\text{\tiny{DM0}}}{d\ln q}, \nonumber \\
		\dot{\Psi}_1&=& \frac{q{\bf K}}{3\varepsilon_\text{\tiny{DM}}}(\Psi_0-2\Psi_2)+\frac{\varepsilon_\text{\tiny{DM}}\,K}{3q}\psi \frac{d\ln f_\text{\tiny{DM0}}}{d\ln q},\nonumber \\
		\dot{\Psi}_l &=& \frac{q{\bf K}}{(2l+1)\varepsilon_\text{\tiny{DM}}}(l\Psi_{l-1}-(l+1)\Psi_{l+1}),\,\,\,\,\,\,l\geq2.
	\end{eqnarray}
	Therefore, by inserting the initial condition, one can solve the above evolution equations numerically.
	Meanwhile, the time averaged  value of the perturbations $\beta_\text{\tiny{DM}}=\frac{\delta p_\text{\tiny{DM}}}{\bar\rho_\text{\tiny{DM}}}-(\frac{\bar\rho_\text{\tiny{DM}} +\bar{P}_\text{\tiny{DM}}}{\bar\rho_\text{\tiny{DM}}})\sigma$ from the last scattering up to today can be estimated as the order of matter anisotropy $\bar\beta_\text{\tiny{DM}}\le 10^{-4}$. \par
\begin{figure}
		\centering
		\includegraphics[width=3in]{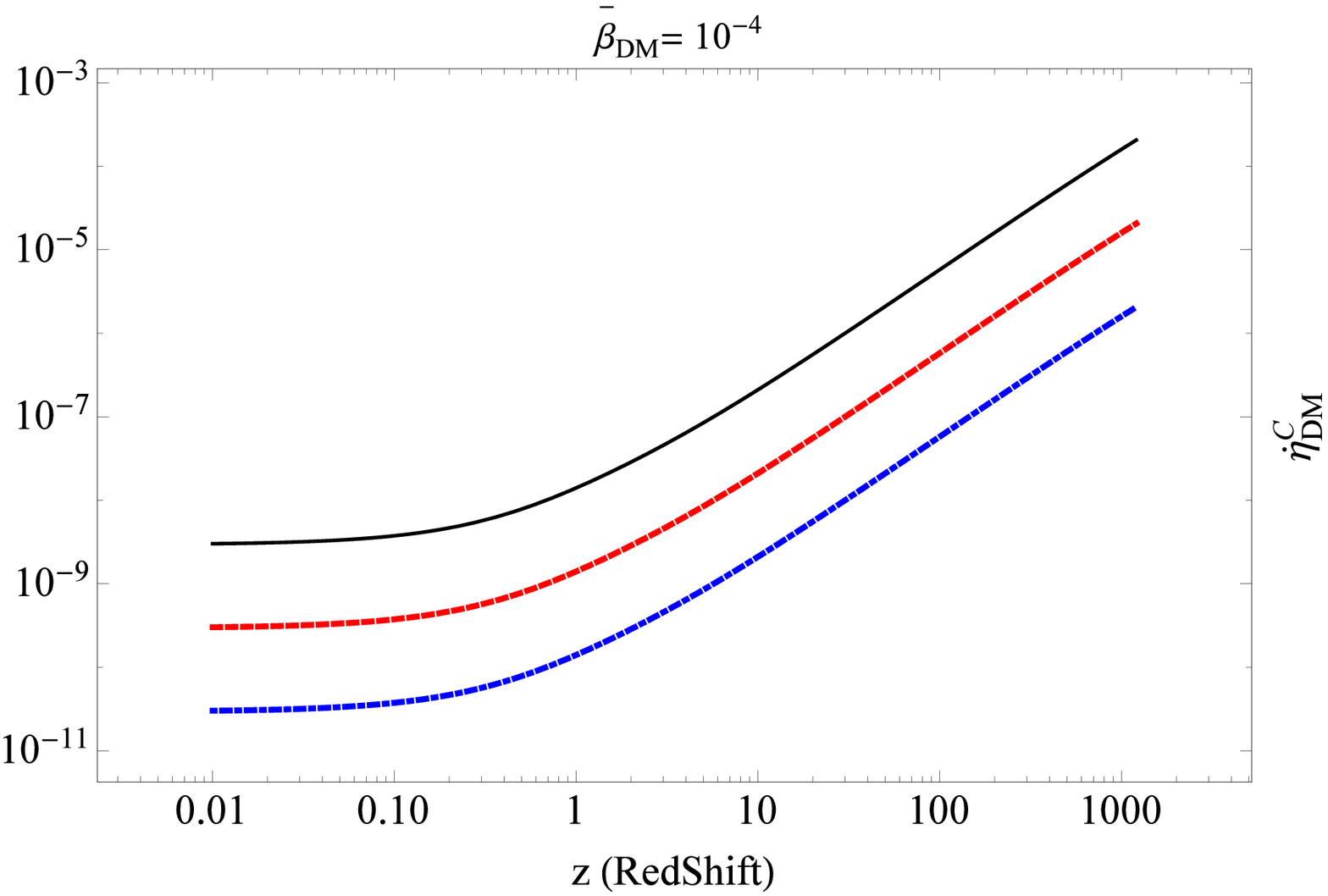}
		\includegraphics[width=3in]{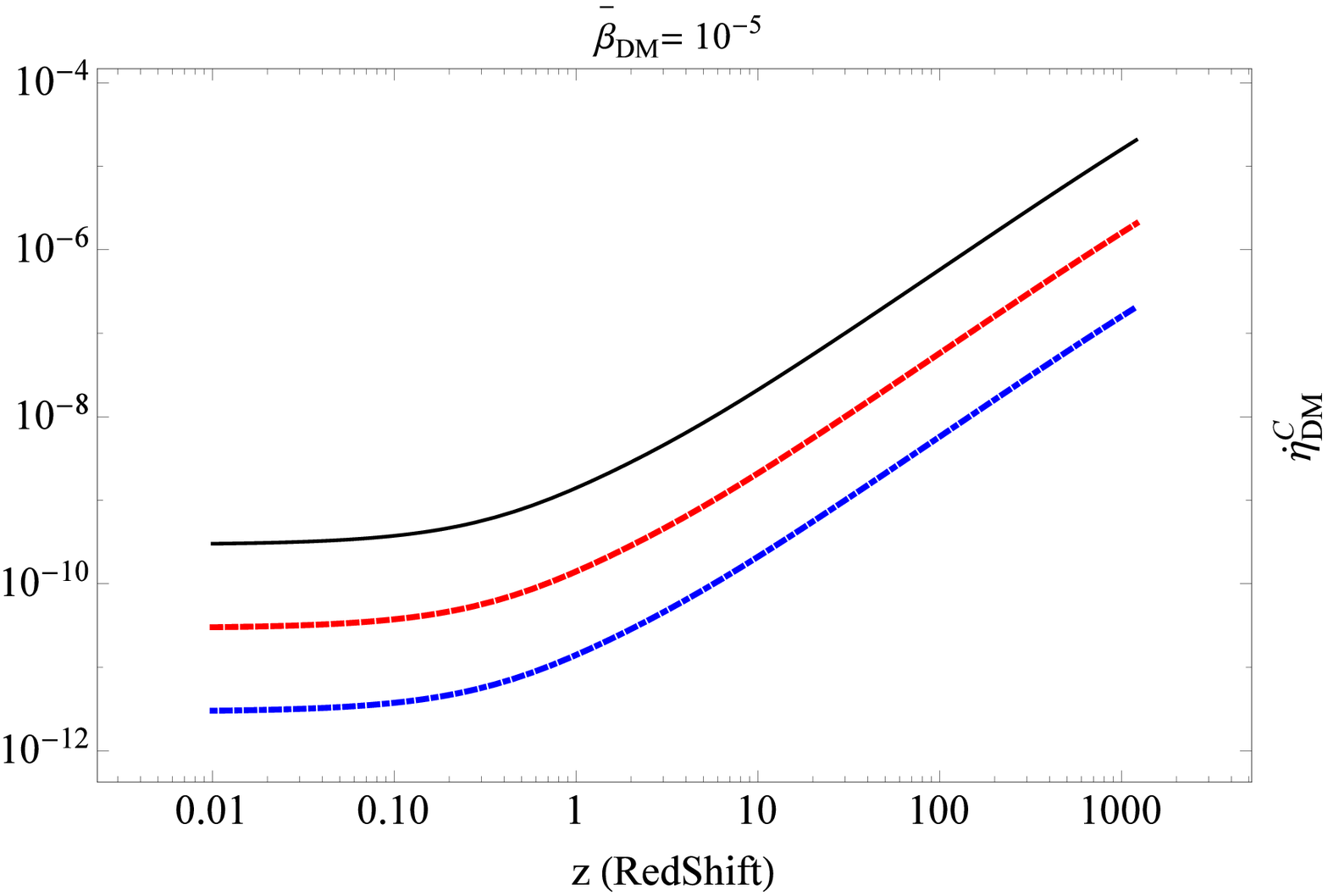}
		\caption{The $\dot{\eta}^\text{\tiny{C}}_\text{\tiny{DM}}$ is plotted as a function of red-shift for different values of mixing  $\theta^2=10^{-3}$ (blue), $\theta^2=10^{-2}$ (red) and $\theta^2=10^{-1}$ (black) with $\bar{\beta}_\text{\tiny{DM}}=10^{-4}$ (left) and $\bar{\beta}_\text{\tiny{DM}}=10^{-5}$ (right). } \label{etadot0}
	\end{figure}

  
	To illustrate how the $\dot{\eta}_\text{\tiny{DM}}^\text{\tiny{C}}$ depends on the red-shift, we have shown the $\dot{\eta}_\text{\tiny{DM}}^\text{\tiny{C}}$  as a function of the red-shift in Fig.(\ref{etadot0}). The $\dot{\eta}_\text{\tiny{DM}}^\text{\tiny{C}}$ are shown for three values of $\theta^2=10^{-3}, 10^{-2}, 10^{-1}$ and  for two cases of the  $\bar{\beta}_\text{\tiny{DM}}=10^{-4}$ and $10^{-5}$. As the figure shows the $\dot{\eta}_\text{\tiny{DM}}^\text{\tiny{C}}$ increases smoothly with red-shift up to $z=1$ then it increases rapidly at high red-shifts.  In fact, the $\dot{\eta}_\text{\tiny{DM}}^\text{\tiny{C}}$ becomes larger for the larger value of the red-shift.
	Furthermore, in the presence of primordial scalar perturbations 
	the CMB temperature and polarization anisotropy are given by the multi-pole moments as follows \cite{Zaldarriaga:1996xe,Seljak:1996is}
	\begin{equation}
		\Delta^{S}_{I,P,V}(\eta,{\bf K},\mu)=\sum_{l=0}^{\infty}(2l+1)(-i)^l\Delta^{l}_{I,P,V}(\eta,{\bf K})P_{l} (\mu),
	\end{equation}
	where $P_{l}(\mu)$ is the Legendre polynomial of rank $l$, 
	and $\mu$ indicates scalar product of the CMB  propagating direction and the wave vector ${\bf K}$. 
	The time derivative in the left side can include the space-time structure and gravitational effects. Besides, the scattering of each plane wave can be described as the transport through a plane parallel medium \cite{{chandra},{kaiser}}. Therefore, the Boltzmann equation for linear and circular polarization (\ref{taupm}) casts into
	\begin{eqnarray}\label{Circ}
		\frac{d}{d\eta}\,\Delta_{V}^{(S)}\,+\,i\,K\,\mu\,\Delta_{V}^{(S)}\,&=&\,C_{e\gamma}^{V}\,+\,\frac{1}{2}\,a(\eta)\,(\dot{\eta}_\text{\tiny{DM}}^{\text{\tiny{C}}-}\,\Delta_{P}^{+\,(S)}\,+\,\dot{\eta}_{\text{\tiny{DM}}}^{\text{\tiny{C}}+}\,\Delta_{P}^{-\,(S)}\,),\nonumber \\
		\frac{d}{d\eta}\,\Delta_{P}^{\pm(S)}\,+\,i\,K\,\mu\,\Delta_{P}^{\pm(S)}\,&=&\,C_{e\gamma}^{\pm}\,\mp\,i\,a(\eta)\dot{\eta}^\text{\tiny{P}}\,\Delta_{P}^{\pm}.
	\end{eqnarray}
	However, the value of linear polarization $\Delta p^{\pm(S)}$ and $\Delta V^{(S)}$ in the direction ${\bf \hat{n}}$ and at the present time $\eta_0$  can be obtained by integrating the  Boltzmann equation along the line of sight \cite{Zaldarriaga:1996xe} and over all the Fourier modes ${\bf K}$ as
	\begin{eqnarray}
		\Delta _{P}^{\pm (S)}(\hat{\bf{n}})
		&=&\int d^3 {\bf{K}} \xi({\bf{K}})e^{\pm2i\phi_{K,n}}\Delta _{P}^{\pm (S)}
		({\mathbf{K}},{\mathbf{p}},\eta),\,\,\,\,\,\label{Boltzmann03} \nonumber\\
		\Delta _{V}^{ (S)}(\hat{\bf{n}})
		&=&\int d^3 {\bf{K}} \xi({\bf{K}})\Delta _{V}^{(S)}
		(\mathbf{K},\mathbf{p},\eta),\,\,\,\,\,\label{Boltzmann3}
	\end{eqnarray}
	where $\xi(\bf{K})$ is a random variable using to
	characterize the initial amplitude of the mode, $\phi_{K,n}$ is the angle required to rotate the $\bf{K}$ and $\hat{\bf{n}}$ dependent basis to a fixed frame in the sky. 
	Therefore we obtain
	\begin{equation}
		\Delta _{P}^{\pm (S)}
		(\mathbf{K},\mu,\eta_0)=\int_0^{\eta_0} d\eta\,\dot\eta_{e\gamma}\,
		e^{ix \mu -\eta_{e\gamma}\mp i\,\eta^\text{\tiny{P}}_\text{\tiny{DM}}}\,\,\Big[ {3 \over 4}(1-\mu^2)\Pi(\mathbf{K},\eta)\Big],
	\end{equation}
	and
	\begin{eqnarray}\label{Vmod}
		\Delta_{V}^{(S)}({\bf{K}}, \mu, \eta_{0})\,&=&\,\frac{1}{2}\,\int_{0}^{\eta_{0}}\,d\eta\,\dot{\eta}_{e\gamma}\,e^{i\,x\,\mu\,-\eta_{e\gamma}}\,[ 3\mu\,\Delta_{V}^{(S)}\,+\,(\frac{\dot{\eta}_\text{\tiny{DM}}^{\text{\tiny{C}}-}}{\dot{\eta}_{e\gamma}}\,\Delta_{P}^{+(S)}\,+\,\frac{\dot{\eta}_\text{\tiny{DM}}^{\text{\tiny{C}}+}}{\dot{\eta}_{e\gamma}}\,\Delta_{P}^{-(S)})],\nonumber \\
		&\simeq& \,\frac{1}{2}\,\int_{0}^{\eta_{0}}\,d\eta\,\dot{\eta}_{e\gamma}\,e^{i\,x\,\mu\,-\eta_{e\gamma}}\,[3\mu\,\Delta_{V}^{(S)}\,+\,2\,\eta_\text{\tiny{B}}\,\frac{\dot{\eta}_\text{\tiny{DM}}^{\text{\tiny{C}}}}{\dot{\eta}_{e\gamma}}\,\Delta_{P}^{(S)}],
	\end{eqnarray}
	where $x\,=\,{\bf{K}}\,(\eta_{0}-\eta)$, $\dot{\eta}_{e\gamma}\,=\,n_{e}\,\sigma_{T}\,\chi_{e}$ and $\eta_{e\gamma}\,=\,\int_{\eta}^{\eta_{0}}\,\dot{\eta}_{e\gamma}\,d\eta$ are the differential optical depth and total optical depth due to the Thomson scattering at time $\eta$  with $\chi_{e}$ being the ionization fraction, respectively. Moreover $\Delta p$ is defined as
	\begin{equation}
		\Delta^{ (S)}_{P}({\bf{K}},\mu,\eta)=\dfrac{3}{4}(1-\mu^{2})\int_{0}^{\eta} d\eta\, e^{ix\mu-\eta_{e\gamma}}\,\dot{\eta_{e\gamma}}\,\Pi({\bf{K}},\eta),
	\end{equation}
	with
	\begin{equation}
		\Pi\equiv\Delta^{S_{2}}_{I}+\Delta^{S_{2}}_{P}-\Delta^{S_{\circ}}_{P}.
	\end{equation}
	\par
\begin{figure}
		\centering
		\includegraphics[width=3in]{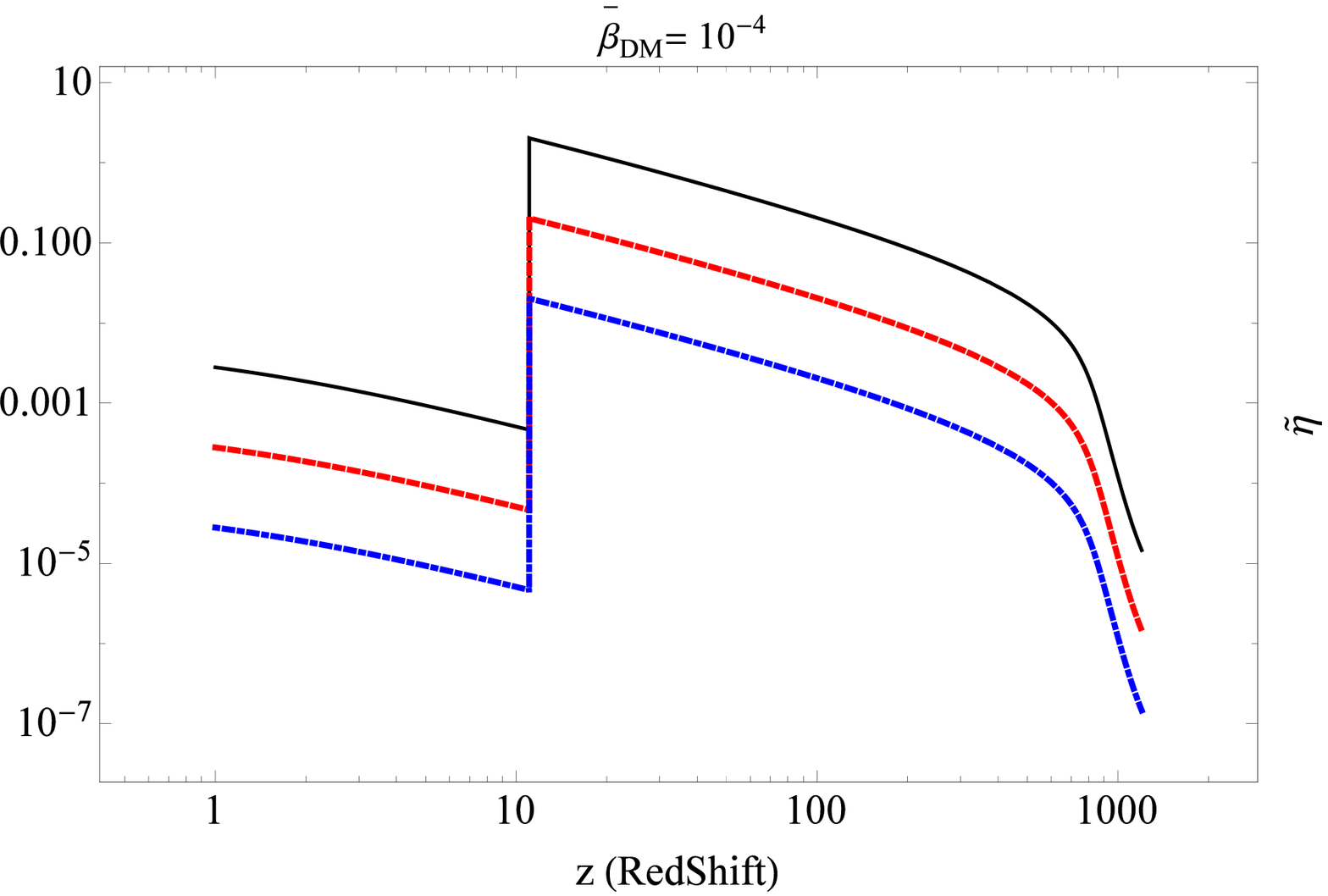}
		\includegraphics[width=3in]{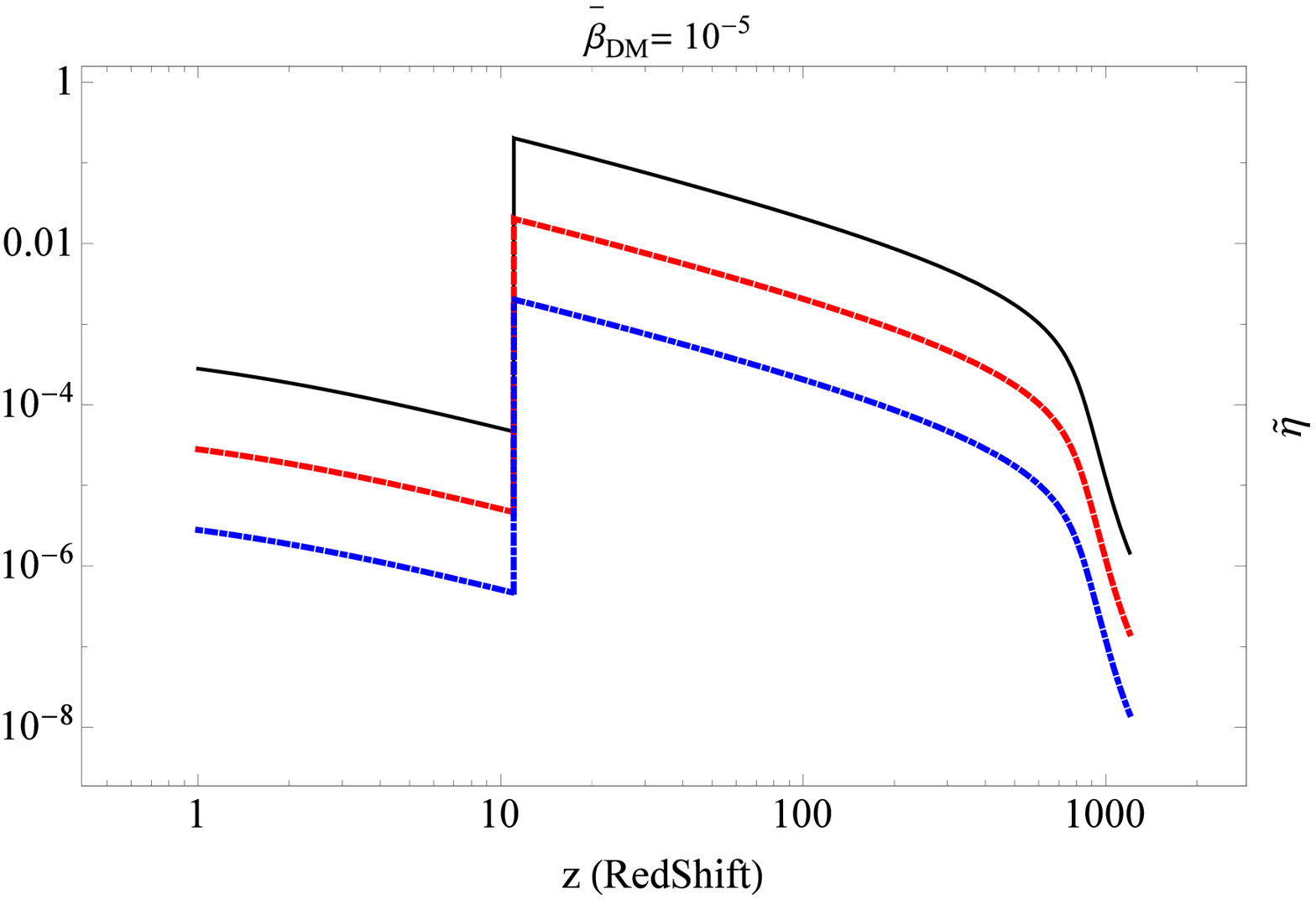}
		\caption{
			The $\tilde{\eta}$ is plotted as a function of red-shift for different values of mixing  $\theta^2=10^{-3}$ (blue), $\theta^2=10^{-2}$ (red) and $\theta^2=10^{-1}$ (black) with $\bar{\beta}_\text{\tiny{DM}}=10^{-4}$ (left) and $\bar{\beta}_\text{\tiny{DM}}=10^{-5}$ (right). }\label{eta0}
	\end{figure}

	Meanwhile, the value of $\frac{\dot{\eta}_\text{\tiny{DM}}^{\text{\tiny{C}}}}{\dot{\eta}_{e\gamma}}$ in (\ref{Vmod}) determines the importance of the CMB-Sterile neutrino  interaction in the CMB polarization and can be obtained as
	\begin{equation}\label{optical}
		\tilde{\eta}\,=\, \frac{\dot{\eta}_\text{\tiny{DM}}^{\text{\tiny{C}}}}{\dot{\eta}_{e\gamma}}\,=\,\frac{\sqrt{2}}{8\pi^{2}}\,\frac{m_{e}^{2}}{\alpha\,\chi_{e}}\,\frac{m_{p}}{p_{0}}\,\frac{\Omega_\text{\tiny{DM}}}{\Omega_\text{\tiny{BM}}}\,G_{\text{\tiny{F}}}\,\theta^{2}\,\bar{\beta}_\text{\tiny{DM}},
	\end{equation}
	where
	$\Omega_\text{\tiny{BM}}=\rho_\text{\tiny{BM}}/\rho_\text{cr}$ and $\Omega_\text{\tiny{DM}}=\rho_\text{\tiny{DM}}/\rho_\text{cr}$ are  the baryonic matter density and the DM mass density parameters, respectively, and  $\rho_\text{cr}$ is the critical density of the universe.
	In the above equation we supposed that the number density of electron or proton is approximately equal to the barionic matter number density $n_e=n_p\simeq n_\text{\tiny{BM}}$. 
	 In Fig.(\ref{eta0}), $\tilde{\eta}$ is plotted as a function of the red-shift for $\theta^2 = 10^{-3}, 10^{-2}$ and $10^{-1}$. The $\tilde{\eta}$ is shown for  $\bar{\beta}_\text{\tiny{DM}}=10^{-4}$ and $10^{-5}$, denoting the importance of the CMB-Sterile neutrino interaction on the CMB circular polarization. It can be seen that the maximum value of $\tilde \eta$ occurs at the red-shift  $z\simeq 10$ and there is a bump at the same red shift. This effect is due to varying  ionized fraction $\chi_e$ around the reionization epoch. As can be seen in Fig .(\ref{eta0}), such a bump always  appears at $z\simeq 10$ for different values of $\bar{\beta}$ and $\theta^2$. 
	In order to study this contribution more accurate, one needs to calculate the total value of the two-point correlation function of the $\Delta V$ mode. To this end, we consider the power spectrum as follows
	\begin{eqnarray}\label{Vmode1}
		C_{V\,l}\,&=&\,\frac{1}{2l+1}\,\sum_{m}\,\langle a^{*}_{V,lm}\,a_{V,lm}\rangle \nonumber\\
		&\simeq&\,\frac{1}{2l+1}\int d^{3}K\,P_{\phi}^{(S)}(K) \left|  \sum_{m}\int d\Omega\,Y_{lm}^{*}({\bf{n}})\,\int_{0}^{\eta_{0}}\,d\eta\,\dot{\eta}_{e\gamma}\,e^{ix\mu\,-\,\eta_{e\gamma}}\,\eta_\text{\tiny{B}}\,\tilde{\eta}\,\Delta_{P}^{(S)} \right| ^{2},
	\end{eqnarray}
	where
	\begin{equation}\label{Vmode2}
		a_{V,lm}\,=\,\int d\Omega\,Y_{lm}^{*}\,\Delta_{V}(\bf{\hat{n}}),
	\end{equation}
	and  the power spectrum $P_{\phi}^{(S)}(K)$ is \cite{earlyuniv} 
	\begin{equation}
		P_{\phi}^{(S)}({\bf{K}})\,\delta({\bf{K'}}-{\bf{K}})\,=\,\langle \xi({\bf{K}})(\xi({\bf{K'}}))\rangle.
	\end{equation}
	Therefore, the above relation leads to the following estimation for $C_{V\,l}$ in terms of the linearly polarized power spectrum as follows
	\begin{equation}\label{cv:new}
		C_{Vl}\, \le \,\tilde{\eta}_\text{\tiny{ave}}^2\, C_{Pl},
	\end{equation}
	where the $\tilde{\eta}_{\text{\tiny{ave}}}$ is the average of $\tilde{\eta}$ in (\ref{optical}) which is calculated for $\theta^2=10^{-2}$ and $\bar{\beta}_\text{\tiny{DM}}=10^{-5}$  as follows

	\begin{eqnarray}\label{eta:0}
	\tilde{\eta}_{\text{\tiny{ave}}}= \frac{1}{\eta_0-\eta_{lss}}(\int_{\eta_0}^{\eta_\text{rei}}\,d\eta\,\tilde{\eta}+\int_{\eta_\text{rei}}^{\eta_{lss}}\,d\eta\,\tilde{\eta})&\simeq&\,3\times 10^{-4}\,(\frac{\theta^{2}}{10^{-2}})(\frac{\bar{\beta}_\text{\tiny{DM}}}{10^{-5}}).
	\end{eqnarray}
	In  (\ref{eta:0}) we have performed the integrals in two regions: I) from the last scattering  to the reionization (recombination) and II) from the reionization to the present. The contribution of the first region (second term in Eq (\ref{eta:0})) is $0.3$ while for the second region (first term in Eq (\ref{eta:0})) one finds a vanishing  contribution as $7\times 10^{-6}$. In fact,  the CMB-Sterile neutrino forward scattering has been produced the main circular polarization contribution during the CMB photon propagation through the recombination region.  
	However, the value of circular power spectrum for $C_{Pl}\sim 0.1 (\mu K)^2 $ will be estimated as
	\begin{equation}\label{cvl}
		C_{Vl}\, \le \,0.01\, (nK)^{2}\,(\frac{\theta^{2}}{10^{-2}})^2(\frac{\bar{\beta}_\text{\tiny{DM}}}{10^{-4}})^{2}.
	\end{equation}
	From the experimental point of view, the upper limit at $95\% $ C.L. on the circular polarization detection ($\ell (\ell+1)C^{VV}_\ell/(2\pi)$) with the 40 GHz polarimeter of Cosmology Large Angular Scale Surveyor (CLASS) has been reported from $0.4 \mu K^2$ to  $13.5 \mu k^2$ between  $1\leq \ell\leq 120$     \cite{Padilla:2019dhz}. The obtained constrained in Eq. \ref{cvl} would be comparable to experimental results by improving the sensitivity of experiments in the near future.  
	\section{Summary and conclusion}\label{sectionV}
	In this paper, we have introduced a new way to examine indirectly the DM signatures.  We have considered a right-handed  neutrino as a preferred WDM candidate which can be coupled to the SM particles within the context of: I- the seesaw type I model (\ref{SN}) and II-the right handed current model (\ref{rhc}). In the second model we have calculated the $W$-boson decay rate to find an upper bound on the coupling constant as $g^2_\text{\tiny{R}}\lesssim 10^{-2}$  
	which is in the same range of acceptable values for $\theta^2$ the coupling constant in the first model.  As the photon-neutrino in both model have the same structures, both models lead to equivalent results for polarization and can not be distinguished in this study.  Nevertheless, we only considered the first model to show that the polarization of cosmic photons which are naturally accelerated to high energy or even as a background can undergo a change via the forward scattering from the DM Sterile neutrino. For this purpose, we considered the GRBs, radio frequency radiation and  the CMB as the sources of high and low energy cosmic photons through the formalism of Stokes parameters and Boltzmann equation. We have shown that the linear polarization of GRBs originated from a collapsing  neutron star can be converted to the circular polarization by scattering from the DM surrounding the star.
	We have found  that the Faraday conversion $\Delta \phi_\text{\tiny{FC}}$ of GRB-Sterile neutrino scattering at both
	the prompt emission and afterglow radiation are about $10^{-21}$ radian and $10^{-18} - 10^{-3}$ radian, respectively.
	We have summarized the Faraday conversion $\Delta \phi_\text{\tiny{FC}}$ for $\theta^{2} \sim 10^{-2}-10^{-6}$ in Tab \ref{tabel1}.  As the table shows the conversion for the prompt emission is too small to be detected in this way.  In contrast, for the long wavelength radio-wave there is a chance to detect the Faraday conversion in our model.  One should note that, here only the maximum value of the Faraday conversion phase shift for the GRB-Sterile neutrino scattering using a simple model is estimated.  Nevertheless, in order to calculate the exact value, one should consider a more complicated  model using the distribution of WDM density in the galaxy and determines the direction of GRBs toward the earth. Moreover, we estimated the Faraday conversion phase shift in radio photon-Sterile neutrino forward scattering. We considered some of the astrophysical radio sources inside and in extra galaxy. The results for the same range of mixing angle $\theta^2$ as GRBs are summarized in Tab. \ref{tabel2}. The Faraday conversion phase shift arising from inter-galactic sources are in the range of $\Delta \phi_\text{\tiny{FC}}\simeq 10^{-6}-10^{-11}$ and from the extra-galactic are in the range $ 10^{-6}-10^{-15}$. The current sensitivity on Faraday Conversion is reported by the PVLAS experiment at the order of $10^{-8}$ rad for the radio wavelength \cite{DellaValle:2015xxa}. We have also shown that the V-mode power spectrum of the polarized CMB as the low energy cosmic photons in the presence of the scalar perturbations can be expressed in terms of the linear polarization power spectrum. We have obtained that the V-mode power spectrum of the CMB photons $C_{Vl}$ caused by CMB-DM scattering is proportional to the linear polarization power spectrum $C_{pl}$ and  the mixing angle  where  for $\theta^2\lesssim 10^{-2}$ and $C_{pl}\sim 0.1 (\mu K)^2$   is of the order of $0.01 \,(nK)^2$, see (\ref{cvl}).   In fact, by considering the current sensitivity for the circular polarization at the order of $\mu K^2$ \cite{Padilla:2019dhz}, our result would be in the range of the accuracy of the future experimental sensitivity.\par 
Finally, since  producing any tiny circular polarization of the cosmic photons might be originated from different effects, we would like to compare our results with some other models.  In our model we found that the $V$-mode is linearly proportional to wavelength $\lambda=(1/p_0)$ as is given in (\ref{deltaphi}).  Furthermore, it depends on the density and the bulk velocity of DM.  Therefore, these facts can be used to compare the obtained results with the other models.  For instance,   the circular polarization of cosmic photons might be generated due to the Compton scattering of cosmic photons in a magnetized intergalactic medium within clusters of galaxies. In this effect  the $V$ mode induced by the magnetic field is proportional to the wavelength as $\lambda^{3}=(1/p_0)^3$, see Eq (6) in ref \cite{Bavarsad:2009hm} and Eq (65) in ref.\cite{Cooray:2002nm}).  In fact, the $\lambda$-dependence of the circular polarization leads to  a different spectrum for the $V$-mode arising from the magnetic field comparing to the $V$-mode from the photon-sterile neutrino forward scattering.  Meanwhile, as a model with the same linear $\lambda$-dependence for the circular polarization, one should considere production of the $V$-mode from photon-active neutrino scattering \cite{Mohammadi:2013dea} \cite{Batebi:2016efn}.   However, the spectra of these two effects can be recognized where the local mass density of DM is dominant in comparison with the active neutrino sources.  In fact, the DM in comparison with the active neutrinos from the CNB (cosmic neutrino background) has a relative global energy density and  bulk velocity  as $\frac{\bar{\rho}_\text{DM}}{\bar{\rho}_\text{CNB}}=\frac{\rho_\text{DM} v_\text{DM}}{\bar{\rho}_\text{CNB}} \simeq 10^{2}$  and $\frac{v_\text{DM}}{v_\text{CNB}}=10^2$, respectively.  Therefore, one expects the induced Faraday conversion due to the GRB/radio photon-DM would be equal or dominant in all electromagnetic wavelength compared to the GRB-CNB scattering.

	\section{Acknowledgment}
	S.Tizchang would like to thank F. Elahi for fruitful discussions.
	\appendix \section{ polarized radiative transfer equation in Stokes-parameter representation}\label{appendix1}
	
In this Appendix, we briefly introduce the polarized radiative transfer equation and show how the Faraday rotation and conversion can be obtained from polarized radiative equation. It is shown that the polarization of a linearly the polarized light which propagates through a medium can be changed as $(Q\leftrightarrow U)$ that is known as Faraday rotation. Meanwhile, Faraday conversion describes the inter-conversion between the linear and circular polarization of the radiation $(Q\leftrightarrow V, U\leftrightarrow V )$. Generally, the polarized radiative transfer equation for a weakly anisotropic medium or homogeneous medium can be expressed as \cite{{Sazonov},{Sazonov2},{Pacholczyk},{Chan:2019umj},{Huang:2011ts}}:
	\begin{equation}\label{AI1}
	\frac{\rm dI_i}{{\rm d} s} =-\kappa_{ij} I_j+\epsilon_i,
	\end{equation}
	where $s$ is the path length of the radiation or equivalently the time that photon is passing through the medium $\rm ds=c\rm dt$ with $c$ being the light velocity which is equal to unity in the natural unit and $I_j$ for $j$  running from 1 to 4 represent the Stokes vector components given by $[I,Q,U,V]$.  Meanwhile, the coefficient tensor $\kappa_{ij}$ denotes the amount of rotation ($f$), conversion ($h,g$), absorption ($\kappa,q,u,\nu$) and $\epsilon_i$ shows the spontaneous emission coefficient.  Therefore, (\ref{AI1}) can be written in a matrix form  as
	\begin{equation} 
	\frac{\rm d}{{\rm d} s}   
	\begin{bmatrix} I \\ Q \\ U \\ V \end{bmatrix} 
	=  
	-    \begin{bmatrix} \kappa & q & u & v \\ q &\kappa & f & -g \\ u & -f & \kappa&h  \\ v& g & -h& \kappa  \end{bmatrix} 
	\begin{bmatrix} I \\ Q \\ U \\ V \end{bmatrix} 
	+   \begin{bmatrix}
	\epsilon_I \\ \epsilon_Q \\ \epsilon_U \\ \epsilon_V 
	\end{bmatrix}.
	\label{eq-PRT}
	\end{equation}
	Howevere, in a  medium without any emission and absorption  $i.e.$ $\kappa=q=u=\nu=0$, and $\epsilon_I =\epsilon_Q = \epsilon_U=\epsilon_V=0$, the polarized radiative transfer equation  result in ${\rm d}I /{\rm d}s = 0$ and reduces  to 
	\begin{equation} 
	\frac{\rm d}{{\rm d} s}   \begin{bmatrix}  Q \\ U \\ V \end{bmatrix}   =  
	-    \begin{bmatrix}    0 & f & -g \\   -f & 0 &h  \\ g & -h& 0  \end{bmatrix} 
	\begin{bmatrix}  Q \\ U \\ V \end{bmatrix}    \  . 
	\label{eq-PRT_x}
	\end{equation}
	Therefore, the time evolution of  $V$ parameter can be written as 
	\begin{equation}
	\frac{\rm dV}{{\rm d} s}=  h\,U-g\,Q ,\label{faraday}
	\end{equation}
	where $g=\frac{d}{dt}\Delta \phi_\text{\tiny{FC}}|_{\tiny Q}$ and $h=\frac{d}{dt}\Delta \phi_\text{\tiny{FC}}|_{\tiny U}$ are the Faraday conversion phase shifts caused by conversion of the linear $Q$ and $U$ polarization to the circular one, respectively.

	\section{Time evolution of the density matrix components via photon-Sterile neutrino interaction}\label{appendix}
	In this appendix, we calculate the time evolution of the density matrix components due to the forward scattering of the photon-Sterile neutrino interaction. \par
\begin{figure}[tb]
		\center
		\includegraphics[scale=0.62]{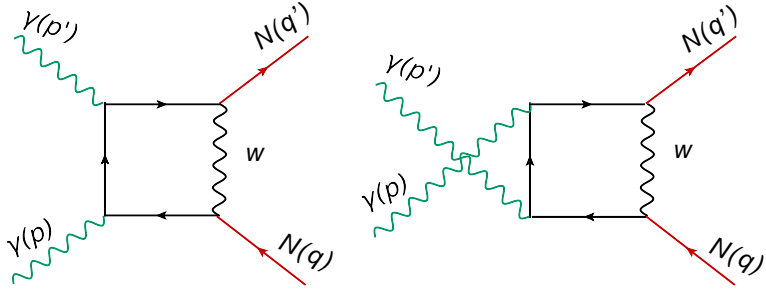}\\
		\caption{The representative Feynman diagrams represent the photon-Sterile neutrino scattering. There are two more Feynman diagrams with antiparticle contributing in the loops.}\label{feyn}
	\end{figure}

  We start with the seesaw Lagrangian given in (\ref{SN}).
	Within the seesaw model, the photon can scatter from Sterile neutrinos at one-loop level with a lepton and weak gauge bosons propagating in the loop. Representative relevant Feynman diagrams are shown in Fig. \ref{feyn}. There are t-channel Feynman diagrams with the loop  involving $W$-boson and the charged leptons (electron, muon and tau ). Furthermore, two additional Feynman diagrams representing the contributions from antiparticles in the loops have been also taken into account. Meanwhile, the 
	contribution from a further s-channel diagram with the $W^+W^-\gamma \gamma$ vertex in which $W$-bosons exchange in a triangle loop as well as  a t-channel diagram similar to Fig \ref{feyn} where three $W$ bosons contribute in a box diagrams  are negligible. \par
	The  electromagnetic free gauge field $A^{\mu}$ and Majorana fermion field $N(x)$, which are self-conjugate, can be indicated as creation $a_{s}^\dagger(p)$ ($b_{r}^\dagger(q)$) and annihilation $a_{s}(p)$ ($b_{r}(q)$) operators for photons (Majorana fermions) as
	\begin{equation}\label{gauge}
		A_{\mu}(x)=\int\frac{d^3 {\bf p}}{(2\pi)^3 2p^0}[a_{s}(p)\epsilon_{s\mu}(p)e^{-ip.x} + a_{s}^\dagger (p)\epsilon_{s\mu}^* (p)e^{ip.x}],
	\end{equation}
	\begin{equation}\label{majorana}
		N(x)= \int \frac{d^3{\bf q} }{(2\pi)^3}\frac{M}{ q^0}\left[ b_r(q) u_{r}(q)
		e^{-iq\cdot x}+ b_r^\dagger (q) v_{r}(q)e^{iq\cdot x}
		\right],
	\end{equation}
	where $\epsilon_{s\mu}(p)$  with $s=1,2$ are the photon polarization 4-vectors of two physical transverse polarization while $u_{r}(q)$ and $v_{r}(q)$ are the Dirac spinors. The creation and annihilation operators respect the following canonical commutation (anti-commutation) relations
	\begin{eqnarray}
		[a_{s}(p),a_{s'}^\dagger (p')]&=&(2\pi)^3 2p^0 \delta_{ss'}\delta^{(3)}({\bf p}-{\bf p'}),\nonumber \\
		\{b_{r}(q),b_{r'}^\dagger (q')\} &=&(2\pi)^3 \frac{q^0}{M} \delta_{rr'} \delta^{(3)}({\bf q}-{\bf q'}).
	\end{eqnarray}
	The leading-order interacting Hamiltonian for this process can be expressed by the scattering amplitude as follows
	\begin{eqnarray}\label{H1}
		H_{I}^0(t)&=&\,\int {d\bf{q}} {d\bf{q'}} {d\bf{p}} {d\bf{p'}}(2\pi)^3 \delta^{(3)} ({\bf{q'}}+{\bf{p'}}-{\bf{q}}-{\bf{p}}) \exp({i[q'^0 +p'^0 -q^0 - p^0]})\nonumber\\
		&\times&[b_{r'}^\dagger({\bf q'}) a_{s'}^\dagger({\bf p'})\mathcal{M}_\text{tot}(N\gamma\,\to\,N\gamma)\, a_{s}({\bf p}) b_{r}({\bf q})],
	\end{eqnarray}
	with ${d \bf{q}} \equiv \frac {d^3 \bf{q}}{(2\pi)^3}\frac{M}{q^0}$, ${d \bf{p}} \equiv \frac {d^3 \bf{p}}{(2\pi)^3}\frac{1}{2 p^0}$ and the total amplitude $\mathcal{M}_\text{tot}$ can be obtained from the sum of all Feynman diagrams in Fig.\ref{feyn}, as follows 
	\begin{eqnarray}\label{m}
		\mathcal{M}_{tot}({\bf q'}r',{\bf p'}s',{\bf q}r,{\bf p}s)&\equiv &\mathcal{M}_{1}({\bf q'}r',{\bf p'}s',{\bf q}r,{\bf p}s) +\mathcal{M}_{2}({\bf q'}r',{\bf p'}s',{\bf q}r,{\bf p}s)\nonumber\\
		&&-\mathcal{M}_{3}({\bf q'}r',{\bf p'}s',{\bf q}r,{\bf p}s) -\mathcal{M}_{4}({\bf q'}r',{\bf p'}s',{\bf q}r,{\bf p}s),
	\end{eqnarray}
	where ${M}_{3,4}({\bf q'}r',{\bf p'}s',{\bf q}r,{\bf p}s)$ are, respectively,  the Hermitian conjugates of ${M}_{1,2}({\bf q'}r',{\bf p'}s',{\bf q}r,{\bf p}s)$ and have been contributed from  antiparticles in the loops as follows
	\begin{eqnarray}\label{m1}
		\mathcal{M}_{1}({\bf q'}r',{\bf p'}s',{\bf q}r,{\bf p}s)&=&\frac{1}{(2\pi)^{4}}\,\frac{e^{2}\,g^{2}}{8}\,\theta^{2}\int d^{4}l\,\, \,\bar{u}_{r^{'}}({\bf q^{'}})\gamma^{\alpha}\,(1-\gamma^{5})\,S_{F}(l+p-p^{'})\slashed{\epsilon}_{s^{'}}({\bf p^{'}})\nonumber\\
		&&S_{F}(p+l)\,\slashed{\epsilon}_{s}({\bf p})\,S_{F}(l)\gamma ^{\beta}\,(1-\gamma^{5})\,u_{r}({\bf q})\,D_{F_{\alpha\beta}}(q-l),
	\end{eqnarray}
	\begin{eqnarray}\label{m2}
		\mathcal{M}_{2}({\bf q'}r',{\bf p'}s',{\bf q}r,{\bf p}s)&=&\frac{1}{(2\pi)^{4}}\,\frac{e^{2}\,g^{2}}{8}\,\theta^{2}\int d^{4}l\,\, \,\bar{u}_{r^{'}}({\bf q^{'}})\gamma^{\alpha}\,(1-\gamma^{5})\,S_{F}(l+p-p^{'})\slashed{\epsilon}_{s}({\bf p})\nonumber\\&&S_{F}(l-p^{'})\,\slashed{\epsilon}_{s^{'}}({\bf q^{'}})\,S_{F}(l)
		\gamma ^{\beta}\,(1-\gamma^{5})\,u_{r}({\bf q})\,D_{F_{\alpha\beta}}(q-l),
	\end{eqnarray}
	\begin{eqnarray}\label{m3}
		\mathcal{M}_{3}({\bf q'}r',{\bf p'}s',{\bf q}r,{\bf p}s)&=&\frac{1}{(2\pi)^{4}}\,\frac{e^{2}\,g^{2}}{8}\,\theta^{2}\int d^{4}l\,\, \,\bar{v}_{r}({\bf q})\gamma^{\alpha}\,(1+\gamma^{5})\,S_{F}(-l)\slashed{\epsilon}_{s}({\bf p})S_{F}(-p-l)\nonumber\\
		&&\,\slashed{\epsilon}_{s^{'}}({\bf q^{'}})\,S_{F}(p^{'}-p-l)\gamma ^{\beta}\,(1+\gamma^{5})\,v_{r}({\bf q'})\,D_{F_{\alpha\beta}}(l-q),
	\end{eqnarray}
	and 
	\begin{eqnarray}\label{m4}
		\mathcal{M}_{4}({\bf q'}r',{\bf p'}s',{\bf q}r,{\bf p}s)&=&\frac{1}{(2\pi)^{4}}\,\frac{e^{2}\,g^{2}}{8}\,\theta^{2}\int d^{4}l\,\, \,\bar{v}_{r}({\bf q})\gamma^{\alpha}\,(1+\gamma^{5})\,S_{F}(-l)\slashed{\epsilon}_{s^{'}}({\bf p^{'}})
		S_{F}(p^{'}-l)\nonumber\\
		&&\,\slashed{\epsilon}_{s}({\bf p})\,S_{F}(p^{'}-p-l)\gamma ^{\beta}\,(1+\gamma^{5})\,v_{r^{'}}({\bf q^{'}})\,D_{F_{\alpha\beta}}(l-q),
	\end{eqnarray}
	where $S_{F}$ is a fermion propagator, the indices $r,r'$ and $s,s'$ denote  the Sterile neutrino and photon spin states, respectively. Moreover, we have considered the contribution of three generations of leptons in each diagram.
	Now to calculate the forward scattering term in (\ref{stoke10}), one should find the commutator $[H_I^0(t),D_{ij}^0({\bf p})]$, then evaluate the expectation value $\langle[H_I^0(t),D_{ij}^0({\bf p})]\rangle$  according to the following operator expectation value
	\begin{equation}
		\langle \, b^\dag_{r'_{i}}(q')b_{r_{j}}(q)\, \rangle
		=(2\pi)^3\delta^3(\mathbf{q}-\mathbf{q'})\delta_{rr'}\delta_{ij}\frac{1}{2}f_\text{\tiny{DM}}(\mathbf{x},\mathbf{q}).
	\end{equation}
	To this end, we substitute (\ref{m}-\ref{m4}) into (\ref{H1}) and then (\ref{stoke10}) and find the time evolution of the density matrix components as
	\begin{eqnarray}\label{rhodott}
		\frac{d}{dt}\rho_{ij}(p)\,&=&\,-\frac{\sqrt{2}}{12\,\pi\,p^{0}}\alpha\,\theta^{2}\,{G}_{\text{\tiny{F}}}\int d{\bf{q}}\,\, (\delta_{is}\rho_{s'j}(p)-\delta_{js'}\rho_{is}(p))\,f_\text{\tiny{DM}}({\bf{x}},{\bf{q}})\,
		\bar{u}_{r}(q)\,\,(1-\gamma^{5})\nonumber\\&&(q\cdot\epsilon_{s}\,\,\slashed{\epsilon}_{s^{'}}\,+\,q\cdot\epsilon_{s^{'}}\,\,\slashed{\epsilon}_{s})\,u_{r}(q)+\frac{\sqrt{2}}{24\,\pi\,p^{0}}\alpha\,\theta^{2}\,{G}_{\text{\tiny{F}}}\int d{\bf{q}}\,\, (\delta_{is}\rho_{s'j}(p)-\delta_{js'}\rho_{is}(p))\,\nonumber\\&&f_\text{\tiny{DM}}({\bf{x}},{\bf{q}})\,    \bar{u}_{r}(q)\,(1-\gamma^{5})\,\slashed{p}\,(\slashed{\epsilon}_{s^{'}}\,\slashed{\epsilon}_{s}\,-\,\slashed{\epsilon}_{s}\,\slashed{\epsilon_{s^{'}}})\,u_{r}(q).
	\end{eqnarray}
	
	
\end{document}